\documentclass[twocolumn,showkeys,prd,nofootinbib,floatfix,preprintnumbers]{revtex4-1}

\usepackage[utf8]{inputenc}
\usepackage{multirow}
\usepackage{amsfonts,amsmath,amssymb} 
\usepackage{graphicx,graphics,color}
\usepackage{gensymb}
\usepackage{longtable}
\usepackage{subfigure}
\usepackage{amsmath}

\usepackage{hyperref}
\usepackage[normalem]{ulem}
\hypersetup{
    colorlinks=true,
    linkcolor=blue}

\usepackage{orcidlink}

\newcommand{\be}{\begin{equation}}
\newcommand{\ee}{\end{equation}}
\newcommand{\ba}{\begin{eqnarray}}
\newcommand{\ea}{\end{eqnarray}}

\begin{document}
\title{Observational constraints on  Diffusion Cosmologies \\
in Unimodular Gravity from DESI DR2} 

\author{Gabriel G\'omez$^1$\,\orcidlink{0000-0002-3618-9824}}
\email{luis.gomezd@umayor.cl}

\author{Guillermo Palma$^2$\,\orcidlink{0000-0001-7326-964X}}
\thanks{corresponding author: guillermo.palma@usach.cl}

\author{Norman Cruz$^{3,4}$\,\orcidlink{0000-0002-0737-3497}}
\email{norman.cruz@usach.cl}

\affiliation{$^1$ Centro Multidisciplinario de F\'isica, Vicerrector\'ia de Investigaci\'on, Universidad Mayor, \\ Camino La Pir\'amide 5750,  Huechuraba, 8580745, Santiago, Chile.\\
$^2$Departamento de F\'{\i}sica, Universidad de Santiago de Chile, Avenida Ecuador 3493, Santiago, Chile,\\ 
$^3$Center for Inter-disciplinary Research in Astrophysics and Space Exploration (CIRAS), Universidad de Santiago de Chile, Av. Libertador Bernardo O’Higgins 3363, Estaci\'on Central, Chile.}

\date{\today} 

\begin{abstract}

Diffusion functions in unimodular gravity induce a dynamical effective cosmological constant, providing an appealing framework in light of recent DESI observations. In this work, we investigate the observational viability of a general class of diffusion models using baryon acoustic oscillation measurements from DESI DR2, Type Ia supernova compilations (Pantheon+, DES-Dovekie, and DESY5), and Cosmic Chronometer data. We find that diffusion models systematically achieve lower best-fit \(\chi^{2}\) values than \(\Lambda\)CDM across all dataset combinations considered, indicating a modest but persistent improvement in goodness of fit. Nevertheless, the reduction in $\chi^{2}$ is insufficient to offset the larger parameter space, leading standard information criteria to favor the simpler $\Lambda$CDM model. Despite this result, all dataset combinations consistently prefer a nonvanishing diffusion contribution corresponding to approximately $20\%$ of the present dark-energy budget, with a posterior probability $P(f_Q>0.05) =96.78\%$. From an observational perspective, diffusion models therefore remain a viable extension of the standard cosmological scenario and motivate the exploration of simpler diffusion parameterizations, particularly in light of upcoming high-precision cosmological surveys.

\end{abstract}

\maketitle

\section{Introduction}
\label{sec:intro}

The standard $\Lambda$CDM model remains the most successful framework for describing a broad range of cosmological observations, with the simplest form of a dark energy (DE) component modeled by the cosmological constant (CC) \cite{Blanchard:2018klb,Efstathiou:2024dvn}. Despite this success, persistent discrepancies between independent cosmological measurements have raised questions about whether the standard framework provides a complete description of the observed Universe. Among these, the Hubble tension is currently one of the most statistically significant \cite{Riess:2021jrx,Knox:2019rjx}. Local measurements give $H_0=73.50\pm0.81,{\rm km\,s^{-1}\,Mpc^{-1}}$, corresponding to a $1.1\%$ precision, while the combination of early-Universe observations with $\Lambda$CDM yields $H_0=67.24\pm0.35,{\rm km\,s^{-1}\,Mpc^{-1}}$, corresponding to a $7.1\sigma$ discrepancy. An independent comparison based on BBN and BAO within flat $\Lambda$CDM, using DESI DR2 data, gives $H_0=68.51\pm0.58,{\rm km\,s^{-1}\,Mpc^{-1}}$ and a $5.0\sigma$ discrepancy with the local determination \cite{H0DN:2025lyy}. These discrepancies directly test the consistency of the standard cosmological framework across cosmic epochs. For recent reviews, see Refs.~\cite{verde2024tale,kamionkowski2023hubble}.

Another important discrepancy is the $\sigma_8$ tension associated with the growth of cosmic structure \cite{Macaulay:2013swa,Battye:2014qga,Alam:2016hwk,Abbott:2017wau}, together with the DESI indication of evolving dark energy. When combined with CMB data and the Pantheon+, Union3, and DESY5 supernova samples, DESI data favor a dynamical dark-energy scenario with $\omega_0>-1$ and $\omega_a<0$ over a pure cosmological constant, with a preference reaching $4.2\sigma$ \cite{DESI:2025zgx}. However, the inclusion of the updated DES-SN5Y sample reduces the preference for departures from $\Lambda$CDM to $3.2\sigma$ \cite{DESI:2025wyn}. This evidence remains under debate, with other analyses finding less compelling support for evolving dark energy \cite{capozziello2025dark,Ong:2025utx,Dinda:2026ktu}.

These persistent observational challenges have motivated the exploration of new physics beyond the standard $\Lambda$CDM framework, including extensions of General Relativity (GR) \cite{cai2026hubble,jia2026review}. Unimodular gravity (UG) provides a particularly interesting setting in this context. Unlike standard GR, UG allows for a non-conservation of the energy-momentum tensor associated with a diffusion process. In particular, a diffusion process that becomes relevant only at late times can provide a natural mechanism for modifying the inferred value of $H_0$ and thereby potentially alleviating the Hubble tension \cite{landau2023cosmological}. Related approaches have addressed the tension through interacting-fluid scenarios \cite{perez2021resolving,LinaresCedeno:2020uxx}, while models based solely on unimodular coordinate transformations have been shown to alleviate the tension at the $2.4\sigma$ level \cite{singh2023unimodular}.

The energy-momentum non-conservation that characterizes UG is encoded in the energy diffusion function (EDF), $Q$, which parametrizes the exchange of energy between the cosmic fluids. Various forms of the EDF have been considered to explore different cosmological scenarios. One physically motivated realization is provided by the continuous spontaneous localization (CSL) framework, in which energy is continuously generated through spontaneous quantum-collapse events~\cite{Pearle:1976ka,Ghirardi:1985mt,Pearle:1988uh,Ghirardi:1989cn}. Other choices of the EDF have instead been motivated primarily by simplicity and phenomenological considerations. For instance, an EDF depending on the jerk parameter was investigated in Refs.~\cite{garcia2019cosmic,garcia2021universe} as a mechanism for generating accelerated expansion within UG. More recently, nonlinear power-law parametrizations in which the EDF depends on the matter density, scalar-field density, or Hubble expansion rate have been constrained using late-time background observations. These analyses indicate a preference for a non-vanishing matter-diffusion interaction over $\Lambda$CDM~\cite{zafari2026testing}.

Recently, a generic reconstruction of the EDF was developed in Ref.~\cite{Gomez:2026eyp} using dynamical-systems techniques. This framework provides a systematic approach to reconstructing and classifying diffusion functions directly from the cosmological phase-space structure, without assuming {\it a priori} a specific phenomenological form. This setup therefore provides the starting point for our analysis. Interestingly, the generic EDF encompasses several diffusion models previously studied in the UG framework. These include the barotropic model, $Q=\alpha\rho$, where $\alpha$ is a constant and $\rho$ is the energy density of the fluid undergoing diffusion~\cite{Corral:2020lxt}, as well as the power-law parametrization $Q(a)=Q_0a^\beta$, where $a$ is the scale factor and $\beta$ is a real parameter~\cite{cruz2026thermodynamicconstraintsfuturesingularities}.

It is worth emphasizing that we do not aim to address or resolve any of the observational tensions discussed above. Instead, we adopt a complementary strategy and focus on the observational viability of the diffusion sector itself. In particular, we address the following questions: Does the data prefer a non-vanishing diffusion contribution? If so, can the observations discriminate between the power-law case and more general forms of diffusion within the generalized framework considered here? And, if diffusion contributes to the present-day dark-energy sector, what fraction of the total dark-energy density can be attributed to it relative to the cosmological-constant contribution? Addressing these questions allows us to assess the viability of diffusion as a component of the cosmological energy budget and its potential implications for dynamical dark energy. Since UG naturally gives rise to a time-dependent dark-energy sector, our results can provide useful indications of dynamical dark energy.

The paper is organized as follows. In Sect.~\ref{sec:model}, we review the fundamental aspects of unimodular gravity relevant to the diffusion framework and present the autonomous dynamical system, which allows a unified description of a broad class of diffusion models. In Sect.~\ref{sec:cosmology}, we confront the diffusion framework with current low-redshift cosmological observations, combining DESI DR2 BAO measurements with complementary probes of the late-time expansion history. We constrain the diffusion sector and its contribution to present-day dark energy, and assess its statistical performance relative to $\Lambda$CDM for different data combinations. Finally, in Sect.~\ref{sec:conclusions}, we summarize the main results and discuss their physical implications.

\section{Diffusion function in Unimodular gravity}
\label{sec:model}

\subsection{Theoretical setup}
\label{sec:theory}

Unimodular gravity follows from the Einstein--Hilbert action by restricting the allowed diffeomorphisms to the volume-preserving subgroup, i.e., by requiring the metric to satisfy $g_{\mu\nu}\,\delta g^{\mu\nu}=0$ under such transformations. The resulting field equations are the trace-free Einstein equations \cite{Einstein:1919gv}:
\begin{equation}\label{TFE}
    R_{\mu\nu}-\frac{1}{4}R\,g_{\mu\nu}=\kappa^2\left(T_{\mu\nu}-\frac{1}{4}T\,g_{\mu\nu}\right),
\end{equation}
where $T$ is the trace of the energy-momentum tensor and $\kappa^2 = 8\pi G_N/c^4$, with $G_N$ Newton's gravitational constant and $c$ the speed of light.
Moreover, within the UG framework, violations of energy-momentum conservation are allowed, provided they are integrable -- that is, if $J_{\mu} \equiv \nabla^{\nu} T_{\nu \mu}$ denotes the nonvanishing violation current, then there exists a scalar $Q$, referred to as the diffusion function (DF), such that $J_{\mu} = \nabla_{\mu} Q$, or equivalently that $J_{\mu}$ is a closed 1-form.

Taking the covariant derivative of Eq.~\eqref{TFE} and using the Bianchi identity $\nabla^{\mu} R_{\mu \nu} = \frac{1}{2}\nabla_{\nu} R$, we obtain
\begin{equation}
    \frac{1}{2} \nabla_{\nu}R = \kappa^2 \left(J_{\nu} -\frac{1}{4} \nabla_{\nu} T \right),
\end{equation}
which integrates directly to give the scalar curvature $R$ as
\begin{equation}
R = 4\kappa^2 \left(Q + \Lambda_{int} - \frac{1}{4} T \right),
\end{equation}
where $\Lambda_{int}$ is an integration constant which, for the particular case $Q=0$, corresponds to Einstein's cosmological constant.
Using this expression allows to recast Eq.~(\ref{TFE}) as 
\begin{equation}\label{UGFE}
    R_{\mu\nu}-\frac{1}{2}R~g_{\mu\nu} + \left(Q + \Lambda_{int} \right) g_{\mu\nu} = \kappa^2 ~T_{\mu\nu}~.
\end{equation}
For the homogeneous, isotropic, and spatially flat Friedmann-Lemaître-Robertson-Walker (FLRW) universe, the field equations take the form
\begin{align}
\label{eq:fried1}
3H^2 &= \kappa^2 \left(\rho_{dm} + \rho_{b} + Q + \Lambda_{0} \right),\\
\label{eq:accel}
2\dot{H} + 3H^2 &= \kappa^2 \left( -p_{dm} - p_{b} + Q + \Lambda_{0} \right),
\end{align}
where an overdot denotes differentiation with respect to cosmic time and $H(t)=\dot{a}/a$ is the Hubble parameter. The subindices $dm$ and $b$ denote dark matter and baryons, respectively. For simplicity, we henceforth denote the integration constant $\Lambda_{int}$ simply by $\Lambda_{0}$, and assume that both dark matter and baryons are pressureless.

\subsection{Dynamical system formulation}
\label{sec:dynamical_systems}

It is convenient to formulate the cosmological equations as an autonomous dynamical system. Besides facilitating the dynamical analysis of the cosmological evolution, this approach provides a unified description of a broad class of diffusion models with a variable diffusion slope, while naturally encompassing several well-known phenomenological diffusion models as limiting cases, as investigated in Ref.~\cite{Gomez:2026eyp}.  Accordingly, we introduce the dimensionless variables

%
\begin{equation}
\Omega_{\rm dm}\equiv\frac{\kappa^{2}\rho_{\rm dm}}{3H^2},
~
\Omega_{b}\equiv\frac{\kappa^{2}\rho_{b}}{3H^2}, ~
\Omega_Q\equiv\frac{\kappa^{2}Q}{3H^2},
~
\Omega_{\Lambda}\equiv\frac{\Lambda_{0}}{3H^2},
\label{eqn:dimensionless_variables}
\end{equation}
which satisfy the Friedmann constraint
\begin{equation}
\Omega_{m}+\Omega_Q+\Omega_{\Lambda
}=1,
\label{eqn:Friedmann_constr}
\end{equation}
where $\Omega_{m}=\Omega_{\rm dm} +\Omega_{b}$ stands for the total matter.
Using the number of e-folds $N=\ln a$ as independent variable, and the acceleration equation,
\begin{equation}
\frac{H'}{H}=-\frac{3}{2}\Omega_m,
\label{H_prime_over_H}
\end{equation}
the cosmological evolution can be recast as the autonomous system
\begin{align}
\Omega_{\rm dm}' &= \lambda(N)\,\Omega_Q +3\Omega_{\rm dm}(\Omega_{m}-1), \label{eqn:Omega_dm_evol}\\
\Omega_{b}' &= 3\Omega_{b}(\Omega_{m}-1), \label{eqn:Omega_b_evol}\\
\Omega_Q' &= -\Omega_Q\left[\lambda(N)-3\Omega_{m}\right], \label{Omega_Q_evol}\\
\lambda' &= \lambda^2\left( 1 - \Gamma_{Q} \right),\label{eqn:evol_slope_diffusion}
\end{align}
where
\begin{equation}
\lambda\equiv-\frac{Q'}{Q},
\qquad
\Gamma_Q\equiv\frac{QQ''}{Q'^2}.
\label{eqn:def_lambda_gamma}
\end{equation}
%
The autonomous character of the system requires $\Gamma_Q$ to be a function of $\lambda$, which in turn implies that $\lambda(N)$ must be locally invertible.

A particularly important case corresponds to
\begin{equation}
\Gamma_Q=1,
\end{equation}
for which $\lambda' =0$ and the diffusion slope remains constant, $\lambda=\beta$. Integrating the definition of $\lambda$ then yields
\begin{equation}
Q(a)=Q_0,a^{-\beta},
\label{power_law}
\end{equation}
corresponding to a power-law diffusion function. Therefore, power-law diffusion models emerge naturally as a fixed-point configuration of the diffusion sector within the phase-space description.

More generally, we consider the \textit{constant-curvature family}
\begin{equation}
\Gamma_Q=\Gamma_{0},
\end{equation}
with $\Gamma_{0}$ constant. In this case, the slope equation can be integrated exactly, giving
\begin{equation}
\lambda(N)=
\frac{\lambda_0}
{1-(1-\Gamma_{0})\lambda_0 N},
\label{sol_slope}
\end{equation}
while the corresponding diffusion function becomes
\begin{equation}
Q(N)=
Q_0
\left[
1-(1-\Gamma_{0})\lambda_0 N
\right]^{\frac{1}{1-\Gamma_{0}}}.
\label{diffusion_Gamma_const}
\end{equation}
It is worth noting that the above expression for $Q(N)$ reduces to the power-law form of Eq.~(\ref{power_law}) in the limiting case $\Gamma_0 \rightarrow 1$, with the identification $\lambda_0 =\beta$. This general class of diffusion models was recently introduced and studied in Ref.~\cite{Gomez:2026eyp}, providing a unified framework that encompasses several diffusion scenarios. Beyond its theoretical appeal, the model opens a direct observational window to investigate whether a dynamical diffusion sector can leave measurable imprints on the late-time expansion history of the Universe. 

An interesting feature of the framework is that the commonly studied power-law diffusion model is recovered in the limit $\Gamma_{0}=1$. Constant-curvature models can thus be interpreted as a natural extension of the power-law scenario, allowing for a time-dependent diffusion slope and, consequently, a richer dynamical behavior of the diffusion sector.

A first inspection of the dynamical system defined by Eqs.~(\ref{eqn:Omega_dm_evol})-(\ref{eqn:evol_slope_diffusion}) shows that the hypersurface $\Omega_{m} + \Omega_Q = 1$ defines an invariant manifold of the dynamics. This follows directly from the evolution equation

\begin{equation}
(\Omega_{m} +\Omega_{Q})' = 3 \Omega_{m} (\Omega_{m} +\Omega_{Q} -1 ), 
\label{eqn:Omega_m_Omega_Q_evol}    
\end{equation}
whose right-hand side vanishes identically on this hypersurface for arbitrary diffusion slope $\lambda$. The condition $\Omega_{\Lambda} = 0$ further restricts the physically relevant phase space via the Friedmann constraint, selecting the invariant submanifold on which the subsequent analysis is carried out.

Moreover, the evolution equation for the diffusion slope $\lambda$ decouples from $(\Omega_{m}, \Omega_{Q}, \Omega_{\Lambda})$ -- that is, it depends only on $\lambda$ itself -- revealing a hierarchical structure in which the diffusion slope evolves independently while still driving the matter and diffusion density parameters through the remaining equations. Within this submanifold, the matter and diffusion sectors exhibit coupled dynamics, a feature that persists for general diffusion functions. This structure accommodates the effective diffusion function defined by Eq.~(\ref{diffusion_Gamma_const}) as a well-posed sector of the model, making it directly amenable to confrontation with cosmological data.

An immediate question that arises is whether current observations exhibit a preference for the power-law case ($\Gamma_{0}=1$) or instead favor a more general description of the diffusion sector. 

We address this question together with the related issue of determining the present-day contribution of diffusion to the dark-energy budget, assuming the decomposition
\begin{equation}
\Omega_{\rm DE}=\Omega_{\Lambda}+\Omega_Q.
\end{equation}
Since $\Omega_{\Lambda}$ originates as an integration constant rather than a fundamental cosmological constant, one could equally adopt the choice $\Omega_{\Lambda}=0$. We nevertheless retain a finite $\Omega_{\Lambda}$ because this choice naturally embeds the $\Lambda$CDM model as a limiting case of the diffusion framework. In particular, when the diffusion component vanishes, $\Omega_{Q}=0$, the integration constant is directly identified with the cosmological constant, and the standard $\Lambda$CDM cosmology is exactly recovered.

This is not the only way to obtain a late-time de Sitter phase. For instance, one may instead choose $\Omega_{\Lambda}=0$ while allowing the diffusion component to approach a constant value $\Omega_{Q}=Q_{0}$. Although both constructions are dynamically viable, retaining a finite integration constant provides a more general parametrization in which the relative contributions of the cosmological-constant and diffusion sectors are left as an observational question rather than fixed a priori.

Although diffusion interacts with the dark-matter sector, its contribution to the background dynamics becomes progressively suppressed toward high redshifts, where the matter density dominates the cosmic energy budget. In this regime, the diffusion sector represents only a negligible fraction of the total energy density, so that the energy transfer from the diffusion sector to dark matter has an insignificant impact on the expansion history. As the Universe evolves toward the dark-energy-dominated epoch, however, the diffusion contribution becomes increasingly relevant and manifests itself as an effective dynamical dark-energy component. Consequently, low-redshift cosmological observations provide the most suitable framework for constraining the diffusion sector and testing possible departures from the standard $\Lambda$CDM cosmology.

\section{Observational Constraints} \label{sec:cosmology}

To constrain the diffusion cosmology model and assess its statistical performance relative to the standard $\Lambda$CDM scenario, we performed a Bayesian parameter estimation analysis using the publicly available package \texttt{Cobaya}~\cite{Torrado:2020dgo}. The exploration of the parameter space was carried out through Markov Chain Monte Carlo (MCMC) sampling, adopting broad priors on all free parameters. For each dataset combination, the chains were evolved until convergence according to the Gelman--Rubin criterion \cite{Gelman:1992zz}, while the resulting posterior distributions and confidence regions were analyzed using \texttt{GetDist}~\cite{Lewis:2019xzd}.

The theoretical predictions entering the likelihood evaluation were generated through a custom implementation of the cosmological background evolution using the publicly available background solver \texttt{CosmoDS} \cite{Roy:2026icy}. This module numerically integrates the autonomous dynamical system introduced in Sec.~\ref{sec:model} and computes the relevant observables, including the Hubble expansion rate, luminosity and angular-diameter distances, and the sound horizon at the drag epoch required by the different likelihoods. These quantities are obtained following the standard cosmological distance formalism (see, e.g., Ref.~\cite{Hogg:1999ad,Weinberg:2013agg}). Throughout the integration, the Friedmann constraint is continuously monitored as a consistency check and is satisfied at the level of machine precision over the entire redshift range considered, ensuring the numerical robustness of the background solutions. The implementation was interfaced with \texttt{Cobaya}, allowing a direct comparison of the diffusion model with current cosmological observations.

\subsection{Observational data and statistical analysis}

The analysis was carried out progressively by incorporating different cosmological datasets. The baseline dataset consists of the recent DESI DR2 BAO compilation \cite{DESI:2025zgx}, which provides twelve independent distance measurements covering the redshift range $0.295 \lesssim z \lesssim 2.33$. We further combined these observations with three independent Type Ia supernova datasets: DESY5~\cite{DES:2024jxu}, the recently recalibrated DES-Dovekie sample~\cite{Popovic:2025glk,DES:2025sig}, and Pantheon+~\cite{Pan-STARRS1:2017jku}, allowing us to investigate the robustness of the diffusion sector against different calibrations and supernova systematics. The redshift ranges covered by these compilations are $0.01<z<2.26$ for Pantheon+ and $0.025<z<1.14$ for DES-Dovekie.

Finally, we extended the analysis by including cosmic chronometer (CC) measurements, which provide direct estimates of the Hubble expansion rate $H(z)$ obtained from differential galaxy ages. The CC compilation employed in this work consists of 30 independent measurements taken from the dataset assembled by Moresco \emph{et al.} \cite{Moresco:2012jh}.

To evaluate the relative performance of the models, we computed the best-fit chi-square statistic
\begin{equation}
\chi^2 = -2\ln \mathcal{L}_{\rm max},
\end{equation}
where $\mathcal{L}_{\rm max}$ denotes the maximum likelihood obtained from the MCMC chains. Although likelihood analyses were performed separately for each observational probe, in what follows we report only the results of the joint analyses. Assuming statistical independence among the datasets, the total likelihood is given by the product of the individual likelihoods,
\begin{equation}
\mathcal{L}_{\rm tot}
=
\prod_i \mathcal{L}_i ,
\end{equation}
which implies
\begin{equation}
\chi^2_{\rm tot}
=
-2\ln \mathcal{L}_{\rm tot}
=
\sum_i \chi_i^2 .
\end{equation}
Consequently, $\chi_{\rm tot}^2$ values correspond to the total goodness-of-fit obtained from the combined datasets. The difference
\begin{equation}
\Delta\chi_{\rm tot}^2
=
\chi^2_{\rm diff}
-
\chi^2_{\Lambda{\rm CDM}},
\end{equation}
was used to quantify the relative goodness of fit between the diffusion model and the standard cosmological scenario.  Negative values of $\Delta\chi_{\rm tot}^2$ indicate that the diffusion model provides a better fit to the data than $\Lambda$CDM.

Because the diffusion scenario introduces additional free parameters, model selection cannot rely solely on improvements in $\chi^2$. We therefore complemented the analysis with the Akaike Information Criterion (AIC) \cite{1974ITAC...19..716A} and the Bayesian Information Criterion (BIC) \cite{Schwarz:1978tpv}, defined respectively as
\begin{equation}
   {\rm AIC} = 
\chi_{\rm tot}^2
+ 2k, 
\end{equation}
and
\begin{equation}
    {\rm BIC} = \chi_{\rm tot}^2 + k\ln N,
\end{equation}
where $k$ denotes the number of free parameters and $N$ is the total number of observational data points included in the fit. The corresponding differences, $
\Delta{\rm AIC}= {\rm AIC}_{\rm diff} - 
{\rm AIC}_{\Lambda{\rm CDM}}$ and $\Delta{\rm BIC}= {\rm BIC}_{\rm diff} - {\rm BIC}_{\Lambda{\rm CDM}}$, 
provide a quantitative measure of whether the improvement in fit quality is sufficient to justify the enlarged parameter space. Negative values favor the diffusion model, whereas positive values indicate a preference for $\Lambda$CDM after accounting for model complexity \cite{Trotta:2008qt}.

The total number of data points entering the different analyses is summarized below:

\begin{itemize}
\item DESI DR2 BAO only: $N=12$.
\item DESI DR2 + Cosmic Chronometers: $N=42$ ($12$ BAO + $30$ CC).
\item DESI DR2 + Supernova sample: $N=12+N_{\rm SN}$.
\item DESI DR2 + PantheonPlus + CC: $N=12+1701+30=1743$.
\end{itemize}

This strategy allows us to separately assess parameter constraints, goodness of fit, and model-selection statistics, thereby providing a comprehensive evaluation of the observational viability of the diffusion cosmology scenario.

The diffusion sector contains the free parameters
$[
{\Omega_{Q0},\lambda_0,\Gamma_{0}}]$ whose corresponding priors  are summarized in table \ref{tab:priors} along with the standard cosmological parameters. 

It is important to note that the interval $0\leq\Gamma_{0}\leq1$ defines a class of constant-curvature diffusion models for which the analytical solution~\eqref{sol_slope} remains regular throughout the cosmological evolution. Consequently, no additional theoretical constraints on $\lambda_{0}$ are required, allowing it to be sampled independently over a broad prior range.

It is important to emphasize that $\Omega_{\Lambda 0}$ is treated as a derived parameter through the Friedmann constraint, rather than as an independent sampling parameter. Consequently, the broad prior adopted for $\Omega_{Q0}$ is automatically constrained by the requirement that the total matter density remains positive: $\Omega_{m0}=1-\left(\Omega_{\Lambda0}+\Omega_{Q0}\right)>0$. This condition guarantees the physical consistency of the sampled cosmological models while allowing the diffusion component to explore a sufficiently broad region of parameter space.

\subsection{Results}

The cosmological parameter constraints obtained from the Markov Chain Monte Carlo analysis are summarized in Table~\ref{tab:constraints}. Since diffusion parameters exhibit noticeably non-Gaussian posterior distributions, we report the median values together with their corresponding $68\%$ credible intervals rather than the posterior means and standard deviations. This choice provides a more robust characterization of asymmetric or skewed posteriors and avoids potential biases associated with a Gaussian approximation.

For a Gaussian posterior distribution, the median, mean, and maximum-likelihood estimator coincide up to statistical fluctuations, and the $68\%$ credible interval reduces to the familiar $\pm1\sigma$ confidence region. This is particularly relevant for the diffusion-sector parameters $\Omega_{Q0}$, $\lambda_0$, and $\Gamma_{0}$, whose marginalized distributions generally display asymmetric tails and parameter degeneracies.

\begin{figure*}
\centering
\includegraphics[width=0.77\hsize,clip]{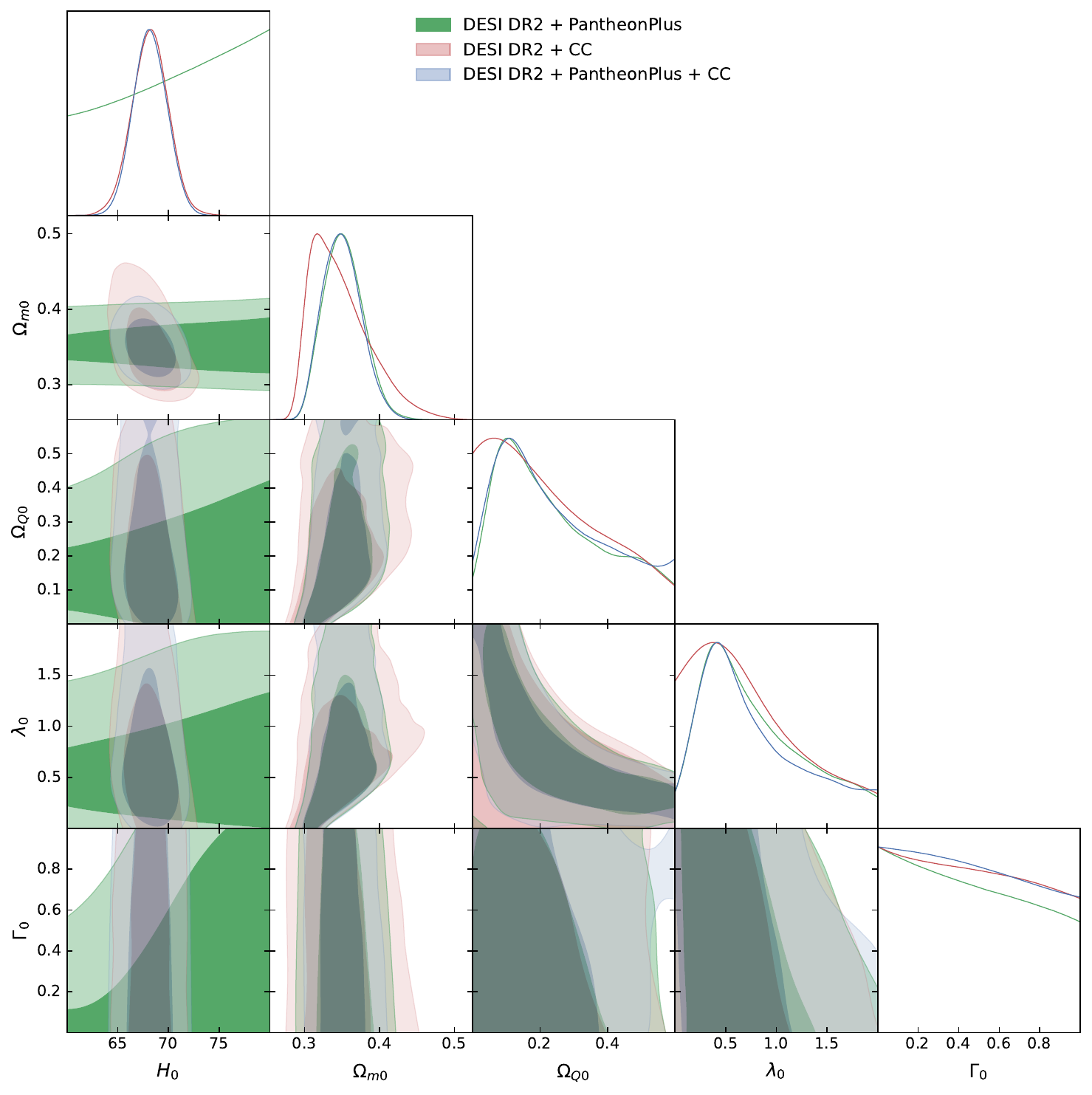}
\caption{
Marginalized posterior distributions and two-dimensional confidence contours ($1\sigma$ and $2\sigma$) for the diffusion-model parameters obtained from the DESI DR2 + PantheonPlus, DESI DR2 + CC and DESI DR2 +  PantheonPlus + CC joint analysis. The figure illustrates the main parameter degeneracies and the broad, non-Gaussian nature of the diffusion-sector constraints. Since the DESI DR2 + DES Y5 and DESI DR2 + DES-Dovekie combinations produce qualitatively similar posterior structures, they are not shown here to avoid redundant information. Their corresponding parameter constraints are summarized in Table~\ref{tab:constraints}.}\label{fig:triangle_contours}
\end{figure*}

Figure~\ref{fig:triangle_contours} shows the  marginalized posterior distributions of the diffusion-sector parameters. In particular, the contours reveal moderate parameter degeneracies, particularly between $\lambda_{0}$ and $\Gamma_{0}$, indicating that different combinations of the diffusion slope and curvature can produce nearly indistinguishable background expansion histories. Consequently, current observations constrain specific parameter combinations more effectively than the individual parameters themselves.

Among the diffusion parameters, $\Omega_{Q0}$ is comparatively well localized, showing that the present-day diffusion density fraction is directly constrained by the data. In contrast, the broader and asymmetric posteriors of $\lambda_{0}$ and $\Gamma_{0}$ indicate that the detailed functional form of the diffusion sector remains only partially determined. This reflects the fact that late-time background observations are primarily sensitive to the integrated expansion history rather than to the underlying diffusion mechanism. Nevertheless, the contours exclude substantial regions of the prior volume and identify a well-defined preferred region of parameter space, demonstrating that the diffusion sector is subject to meaningful observational constraints.

Table~\ref{tab:constraints} reports the marginalized constraints obtained for all observational data combinations considered in this work. For each dataset, we show the median and $68\%$ credible interval of the cosmological parameters in both the diffusion model and the $\Lambda$CDM scenario. The table allows a direct comparison between the standard cosmological parameters, such as $H_0$ and $\Omega_{m0}$, and the additional diffusion-sector quantities. In particular, $\Omega_{Q0}$ quantifies the present-day contribution of the diffusion component, while $\lambda_0$ and $\Gamma_{0}$ characterize the shape and evolution of the diffusion sector. The total dark-energy contribution can be directly inferred from $\Omega_{{\rm DE}0}=\Omega_{\Lambda0}+\Omega_{Q0}=1-\Omega_{m0}$.

\begin{table}[t]
\centering
\caption{
Uniform prior ranges adopted in the Monte Carlo analysis.
The parameters $\Omega_{Q0}$, $\lambda_0$, and $\Gamma_{0}$
are specific to the diffusion cosmology scenario.
}
\begin{tabular}{cc}
\hline\hline
Parameter & Prior \\
\hline

$H_0$ & $[60,80]$ \\

$\Omega_{\rm dm0}$ & $[0.1,0.5]$ \\

$\Omega_{b0}$ & $[0.01,0.01]$ \\

$M_B$ & $[-20,-18]$ \\

$\Omega_{Q0}$ & $[0,0.60]$ \\

$\lambda_0$ & $[0,2]$ \\

$\Gamma_{0}$ & $[0,1]$ \\

\hline\hline
\end{tabular}
\label{tab:priors}
\end{table}

\begin{table*}[t]
\centering
\caption{
Median parameter values and $68\%$ credible intervals obtained for the diffusion model and $\Lambda$CDM using the different dataset combinations. In addition to the standard cosmological parameters, the diffusion model includes the present-day diffusion density fraction $\Omega_{Q0}$ and the parameters $\lambda_0$ and $\Gamma_{0}$. The present dark-energy density satisfies
$\Omega_{{\rm DE}0}=\Omega_{\Lambda0}+\Omega_{Q0}=1-\Omega_{m0}$.}
\begin{tabular}{llcccccc}
\hline\hline
Dataset &
Model &
$H_0 [\mathrm{Km~s^{-1}~Mp^{-1}}]$ &
$\Omega_{m0}$ &
$\Omega_{Q0}$ &
$\lambda_0$ &
$\Gamma_{0}$
\\
\hline

DESI DR2 + DESY5
&
Diffusion
&
$71.8145^{+5.4554}_{-6.8478}$
&
$0.3827^{+0.0266}_{-0.0272}$
&
$0.2776^{+0.1690}_{-0.1257}$
&
$0.8067^{+0.4921}_{-0.3240}$
&
$0.4310^{+0.3378}_{-0.2842}$
\\

&
$\Lambda$CDM
&
$71.2581^{+6.0520}_{-6.1901}$
&
$0.3105^{+0.0082}_{-0.0078}$
&
--
&
--
&
--
\\[0.5ex]

DESI DR2 + DES-Dovekie
&
Diffusion
&
$71.5211^{+5.6007}_{-6.7738}$
&
$0.3536^{+0.0261}_{-0.0242}$

&
$0.2250^{+0.1873}_{-0.1204}$
&
$0.6557^{+0.5466}_{-0.3171}$
&
$0.4345^{+0.3385}_{-0.2886}$
\\

&
$\Lambda$CDM
&
$71.8196^{+5.0939}_{-7.0418}$
&
$0.3062^{+0.0079}_{-0.0073}$
&
--
&
--
&
--
\\[0.5ex]

DESI DR2 + PantheonPlus
&
Diffusion
&
$71.6054^{+5.6319}_{-6.8507}$

&
$0.3496^{+0.0277}_{-0.0252}$
&
$0.2050^{+0.1819}_{-0.1142}$
&
$0.6725^{+0.5513}_{-0.3477}$
&
$0.4406^{+0.3389}_{-0.2914}$

\\

&
$\Lambda$CDM
&
$70.7364^{+5.9616}_{-6.6334}$
&
$0.3047^{+0.0081}_{-0.0076}$
&
--
&
--
&
--
\\[0.5ex]

DESI DR2 + CC
&
Diffusion
&
$68.2105^{+1.7309}_{-1.7581}$
&
$0.3402^{+0.0452}_{-0.0304}$
&
$0.1962^{+0.1854}_{-0.1232}$
&
$0.6134^{+0.5523}_{-0.3651}$
&
$0.4701^{+0.3228}_{-0.3094}$
\\

&
$\Lambda$CDM
&
$69.2792^{+1.7005}_{-1.6171}$
&
$0.2982^{+0.0085}_{-0.0089}$
&
--
&
--
&
--
\\[0.5ex]

DESI DR2 + PantheonPlus + CC
&
Diffusion
&
$68.1793^{+1.7049}_{-1.6543}$
&
$0.3476^{+0.0269}_{-0.0246}$
&
$0.2035^{+0.1854}_{-0.1121}$
&
$0.6366^{+0.5528}_{-0.3261}$
&
$0.4533^{+0.3365}_{-0.2939}$
\\

&
$\Lambda$CDM
&
$68.7961^{+1.7147}_{-1.6239}$
&
$0.3055^{+0.0077}_{-0.0078}$
&
--
&
--
&
--
\\

\hline\hline
\end{tabular}
\label{tab:constraints}
\end{table*}

\begin{table*}[t]
\centering
\caption{Comparison between the diffusion model and $\Lambda$CDM for different combinations of DESI DR2 BAO, Type Ia supernovae, and cosmic chronometer data. Negative values of $\Delta\chi^2$ indicate a better fit of the diffusion model relative to $\Lambda$CDM. The quantities $\Delta{\rm AIC}$ and $\Delta{\rm BIC}$ are defined as the diffusion-model values minus the corresponding $\Lambda$CDM values.}
\begin{tabular}{lcccccc}
\hline\hline
Dataset &
$\Delta\chi^2_{\rm BAO}$ &
$\Delta\chi^2_{\rm SN}$ &
$\Delta\chi^2_{\rm CC}$ &
$\Delta\chi^2_{\rm tot}$ &
$\Delta{\rm AIC}$ &
$\Delta{\rm BIC}$ \\
\hline

DESI DR2 + DESY5
& $-3.24$
& $-8.01$
& ---
& $-11.25$
& $-5.22$
& $+11.34$
\\

DESI DR2 + DES-Dovekie
& $-2.24$
& $-3.67$
& ---
& $-5.91$
& $+0.10$
& $+16.65$
\\

DESI DR2 + PantheonPlus
& $-1.88$
& $-2.77$
& ---
& $-4.66$
& $+1.34$
& $+17.70$
\\

DESI DR2 + CC
& $-1.27$
& ---
& $+0.04$
& $-1.23$
& $+4.77$
& $+10.62$
\\

DESI DR2 + PantheonPlus + CC
& $-1.97$
& $-2.68$
& $+0.14$
& $-4.51$
& $+1.49$
& $+17.90$
\\

\hline\hline
\end{tabular}
\label{tab:model_comparison}
\end{table*}

A few remarks emerge immediately from the model comparison statistics, which are summarized in table \ref{tab:model_comparison} and quantitatively described as follows:
\begin{itemize}

\item The diffusion model consistently improves the fit relative to $\Lambda$CDM for all supernova compilations considered, as indicated by the negative values of $\Delta\chi^2$. The improvement is observed both in the BAO and SN measurements separately, demonstrating that the effect is not driven by a single dataset.

\item The largest improvement is obtained for the DESI DR2 + DESY5 dataset, where the diffusion model reduces the total chi-square by more than eleven units, corresponding to $\Delta\chi^2_{\rm tot}=-11.25$.

\item The inclusion of PantheonPlus and CC measurements does not eliminate the preference for diffusion. In the most comprehensive analysis, DESI DR2 + PantheonPlus + CC, the diffusion model still achieves
$\Delta\chi^2_{\rm tot}=-4.51$,
with the improvement arising primarily from the BAO and supernova sectors, while the CC contribution remains essentially neutral.

\item The information criteria lead to a more conservative conclusion. Since the diffusion scenario introduces three additional parameters, both AIC and, more significantly, BIC penalize the increased model complexity. While $\Delta{\rm AIC}$ indicates that the diffusion model is statistically indistinguishable from $\Lambda$CDM or provides, at most, weak evidence in its favor, $\Delta{\rm BIC}$ strongly disfavors the diffusion scenario due to the stronger penalty associated with the enlarged parameter space and data points. Therefore, although the diffusion model systematically improves the quality of the fit, the corresponding reduction in $\Delta\chi_{\rm tot}^2$ is generally insufficient to compensate for the additional degrees of freedom, particularly when evaluated using the more stringent BIC criterion.

\end{itemize}

\begin{table*}[t]
\centering
\caption{Constraints on the diffusion sector obtained from different combinations of low-redshift cosmological observations. The quantity
$f_Q \equiv \Omega_{Q0}/\Omega_{{\rm DE}0}$
measures the fractional contribution of the diffusion component to the present dark-energy density. The reported values correspond to the posterior median with $68\%$ credible intervals. The last column reports the posterior probability that the diffusion fraction exceeds $5\%$.}
\begin{tabular}{lccc}
\hline\hline
Dataset &
$f_Q$ &
$P(f_Q>0.05)$ \\
\hline

DESI DR2 + DESY5
& $0.453^{+0.281}_{-0.210}$

& $99.60\%$
\\

DESI DR2 + DES-Dovekie
& $0.351^{+0.300}_{-0.191}$
& $98.24\%$
\\

DESI DR2 + PantheonPlus
& $0.317^{+0.289}_{-0.180}$
& $97.06\%$
\\

DESI DR2 + CC
& $0.302^{+0.303}_{-0.194}$
& $93.96\%$
\\

DESI DR2 + PantheonPlus + CC
& $0.315^{+0.293}_{-0.177}$
& $96.78\%$
\\

\hline\hline
\end{tabular}
\label{tab:diffusion_constraints}
\end{table*}

Although the Bayesian Information Criterion consistently favors the simpler $\Lambda$CDM model, particularly for the full joint analysis, Table~\ref{tab:diffusion_constraints} reveals a remarkably stable diffusion sector across all dataset combinations considered. For instance, independent supernova compilations consistently favor a non-vanishing diffusion contribution of order
\[
\Omega_{Q0}\sim 0.2,
\]
corresponding to roughly
\[
f_Q \equiv \frac{\Omega_{Q0}}{\Omega_{{\rm DE}0}}
\sim 0.3-0.4.
\]
The persistence of this preferred region across different datasets suggests that the diffusion component is not merely an artifact of a particular supernova compilation, but rather reflects a robust feature of the parameter space currently allowed by late-time cosmological observations.

An important result is that the inclusion of additional low-redshift observations does not weaken the evidence for diffusion. When cosmic
chronometer measurements are added to the DESI DR2 and PantheonPlus
compilation, the inferred diffusion fraction remains essentially
unchanged,
\begin{equation}
f_Q = 0.315^{+0.293}_{-0.177},
\end{equation}
with a posterior probability
\begin{equation}
P(f_Q>0.05)=96.78\%.
\end{equation}
This value is nearly identical to that obtained from DESI DR2 and
PantheonPlus alone, indicating that the preference for a non-negligible
diffusion sector is not driven by a particular supernova compilation or
by a specific distance indicator.

\begin{figure*}
\centering
\includegraphics[width=0.57\hsize,clip]{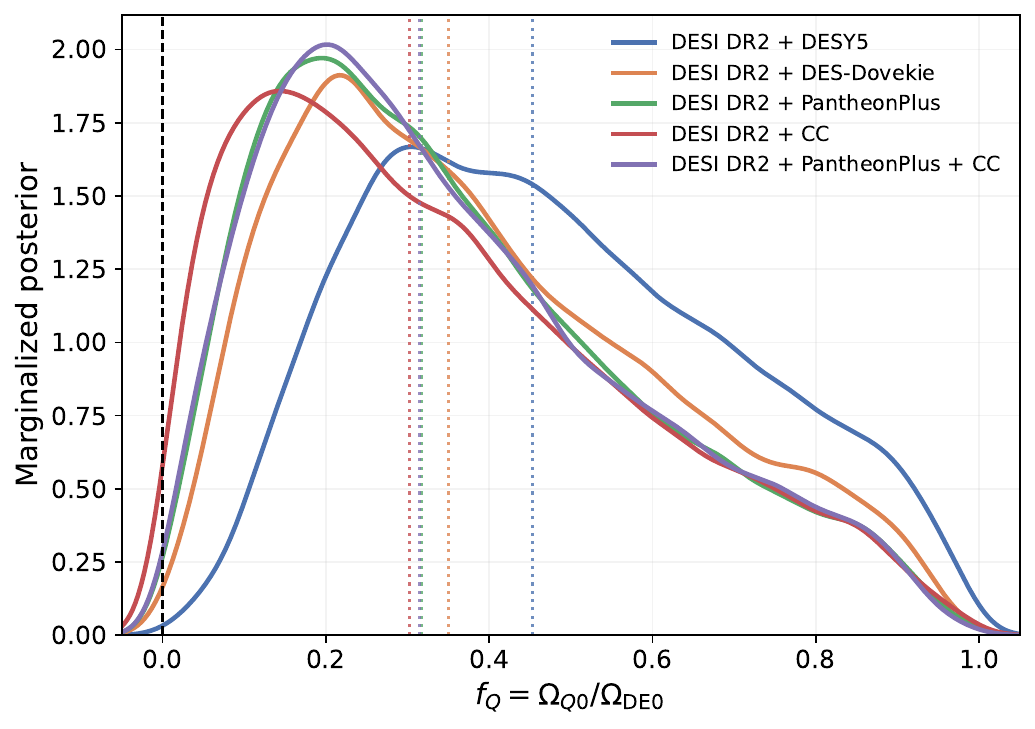} \caption{Marginalized posterior distributions of the diffusion fraction
obtained from different combinations of low-redshift cosmological observations. The posterior distributions are reconstructed from the MCMC samples using a Gaussian kernel density estimator. The vertical dashed lines indicate the corresponding median values reported in Table~\ref{tab:diffusion_constraints}, while the vertical black line denotes the $\Lambda$CDM limit, $f_Q=0$, corresponding to the absence of diffusion. For all dataset combinations, the posterior distributions are shifted toward positive values, favoring a non-vanishing diffusion contribution to the present-day dark-energy sector.} \label{fig:posteior_diffusion}
\end{figure*}

Figure \ref{fig:posteior_diffusion} shows the marginalized posterior distribution of the diffusion fraction $f_{Q}$ for all dataset combinations considered\footnote{The small tails extending into $f_{Q}<0$ are an artifact of the Gaussian kernel density estimator, whose kernels have infinite support. No MCMC samples lie in the unphysical region $f_{Q}<0$.}. The posterior remains centered around $f_{Q}\sim 0.2$ in all cases, with substantial overlap between datasets. In particular, neither the addition of cosmic chronometers nor the inclusion of PantheonPlus data drives the posterior toward $f_{Q}=0$. \textit{This demonstrates that the preference for a nonvanishing diffusion sector is stable against changes in the low-redshift dataset combination and is not driven by any single observational probe}.

\begin{figure*}
\centering
\includegraphics[width=0.57\hsize,clip]{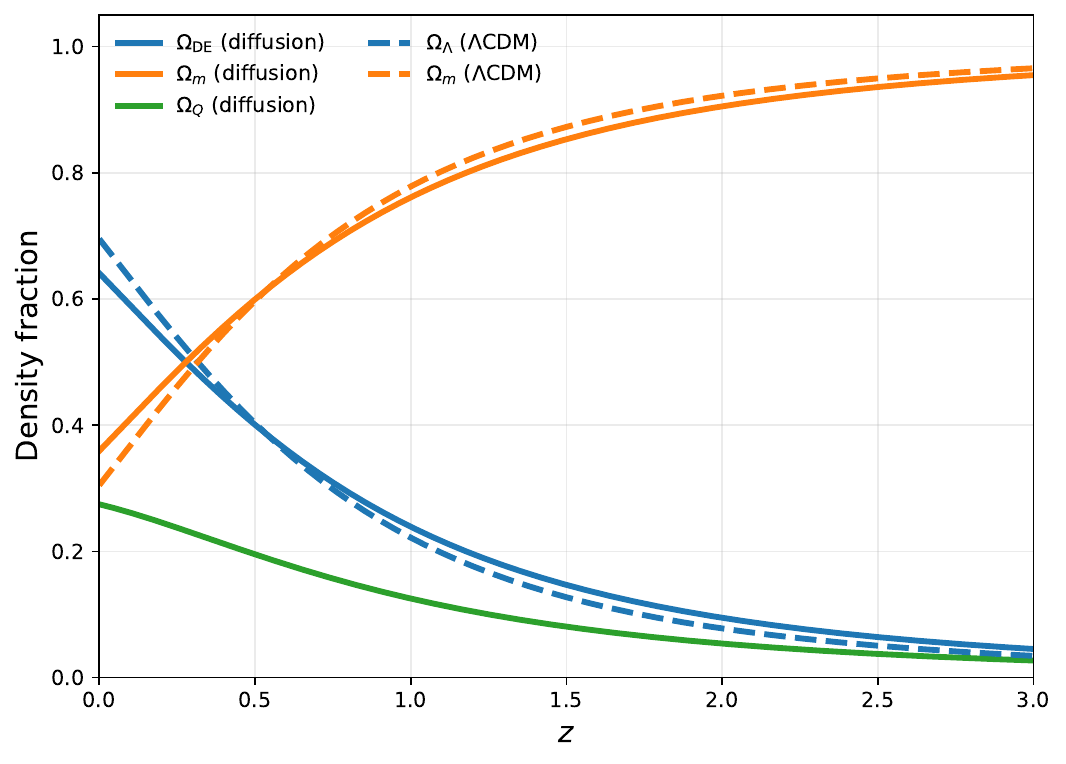}
\caption{
Evolution of the density fractions in the diffusion cosmology model compared with the standard $\Lambda$CDM scenario. The solid curves correspond to the diffusion model, while the dashed curves denote the $\Lambda$CDM prediction. The effective dark-energy contribution $\Omega_{\rm DE}(z)$ is decomposed into the diffusion component $\Omega_Q(z)$ and $\Omega_{\Lambda_{0}}(z)$. The initial conditions for the cosmological parameters are fixed to the best-fit values obtained from the joint DESI DR2 + PantheonPlus + CC analysis. In particular, the present-day dark-energy density is composed of both a cosmological constant-like component and a diffusion component, with the latter contributing approximately $22\%$ of the total dark-energy density.
}\label{fig:Omega_evolution}
\end{figure*}

\subsection{Late-time cosmological evolution}

To illustrate the observational implications of the diffusion scenario, we investigate the late-time cosmological evolution in the redshift interval $0 \leq z \leq 3$, using the best-fit values inferred from the joint DESI DR2 + Pantheon+ + CC analysis. Throughout this section, these values are adopted as initial conditions and compared directly with the corresponding best-fit $\Lambda$CDM model. The corresponding best-fit parameters for the diffusion cosmology are
$H_0=68.104~{\rm km\,s^{-1}\,Mpc^{-1}}$,
$\Omega_{m0}=0.359$,
$\Omega_{Q0}=0.275$,
$\Omega_{\Lambda0}=0.366$,
$\Omega_{{\rm DE}0}=0.641$,
$\lambda_0=0.645$,
For comparison, the best-fit $\Lambda$CDM parameters are
$H_0=68.654~{\rm km\,s^{-1}\,Mpc^{-1}}$,
$\Omega_{m0}=0.305$,
and $\Omega_{\Lambda0}=\Omega_{{\rm DE}0}=0.695$.

Figure~\ref{fig:Omega_evolution} shows the redshift evolution of the fractional energy densities  together with the corresponding $\Lambda$CDM prediction. Solid curves denote the diffusion model, while dashed curves represent the standard $\Lambda$CDM scenario. The quantities displayed are the matter density fraction $\Omega_m(z)$, the total dark-energy fraction $\Omega_{\rm DE}(z)$, and the diffusion component $\Omega_Q(z)$.

Several features emerge from this comparison.

\begin{itemize}

\item
At the present epoch ($z=0$), the diffusion sector contributes a substantial fraction of the total dark-energy budget. Using the best-fit values, one can infer that nearly half of the present dark-energy density is carried by the diffusion component,
\begin{equation}
f_Q \equiv
\frac{\Omega_{Q0}}{\Omega_{{\rm DE}0}}
\simeq 0.43.
\end{equation}

\item The total dark-energy density in the diffusion model remains remarkably close to the $\Lambda$CDM prediction throughout most of the cosmic evolution, particularly in the ranges $0.4\lesssim z\lesssim0.7$ and $z\gtrsim3$. Both models therefore reproduce a late-time accelerated expansion driven by a dominant dark-energy component. At the present epoch, however, noticeable differences emerge: the diffusion model predicts $\Omega_{\rm DE0}\simeq0.64$, compared to $\Omega_{\rm DE0}\simeq0.70$ in $\Lambda$CDM, with the deficit in dark energy compensated by an enhanced dark-matter contribution. Despite these differences, the overall evolution of the density fractions exhibits only mild deviations across the entire redshift range considered. This explains why the diffusion scenario is able to fit current background observations while remaining phenomenologically consistent with the standard cosmological picture.

\item
Although the total dark-energy sector closely mimics $\Lambda$CDM, its internal composition is fundamentally different. Instead of being entirely described by a cosmological constant, the dark-energy budget is dynamically split into a cosmological-constant-like contribution and a nonvanishing diffusion component. Consequently, the diffusion model introduces additional internal dynamics without substantially altering the overall expansion history.

\item The dark matter fraction also exhibits a small departure from the standard $\Lambda$CDM evolution. While both models follow nearly identical trajectories over most of the redshift range, the diffusion scenario predicts a marginally later onset of dark-energy domination. This delay is very small and remains consistent with the percent-level deviations observed in the expansion history. This behavior originates from the continuous energy exchange between the dark-matter and the diffusion sector.

\item
At redshifts $z\gtrsim1$, the diffusion component becomes progressively less important and the Universe approaches the standard matter-dominated regime. This ensures compatibility with observational data at intermediate and high redshift. 

\item
The persistence of a sizable diffusion contribution at late times is particularly noteworthy because it survives the successive inclusion of independent cosmological probes. The combined DESI DR2 + PantheonPlus + CC analysis continues to favor a non-negligible diffusion fraction, indicating that the diffusion sector is not driven to zero by current background observations.

\end{itemize}

Overall, Fig.~\ref{fig:Omega_evolution} illustrates that the diffusion cosmology does not simply reproduce $\Lambda$CDM. Instead, it provides a distinct physical realization of the dark-energy sector in which a substantial diffusion component remains present at late times while preserving the successful background expansion history required by current observations.

\begin{figure*}
\centering
\includegraphics[width=0.57\hsize,clip]{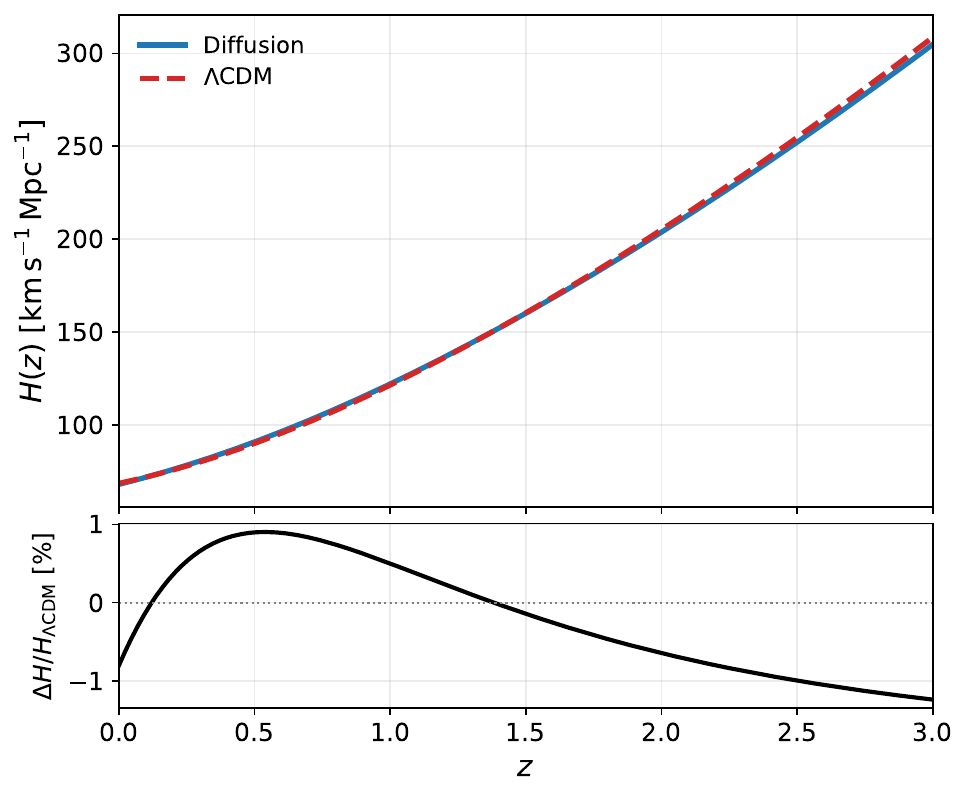}
\caption{
Comparison of the background expansion history predicted by the diffusion cosmology model and the $\Lambda$CDM model using the best-fit parameters obtained from the joint DESI DR2 + PantheonPlus + CC analysis. The upper panel shows the Hubble rate $H(z)$ for both models, while the lower panel displays the fractional deviation
$\Delta H/H_{\Lambda{\rm CDM}} \equiv (H_{\rm diff}-H_{\Lambda{\rm CDM}})/H_{\Lambda{\rm CDM}}$
expressed as a percentage. The diffusion model remains extremely close to the standard cosmological scenario over the entire redshift interval $0\le z\le3$, with deviations below the percent level and a maximum departure around $z\sim0.5$. Positive (negative) values indicate a faster (slower) expansion rate relative to $\Lambda$CDM.} \label{fig:Hubble_comparison}
\end{figure*}

Figure~\ref{fig:Hubble_comparison} compares the background expansion history predicted by the diffusion cosmology model against the best-fit $\Lambda$CDM solution obtained from the same DESI DR2 + PantheonPlus + CC dataset. The upper panel displays the Hubble rate $H(z)$, while the lower panel shows the relative deviation
\begin{equation}
\frac{\Delta H}{H_{\Lambda{\rm CDM}}}
=
\frac{H_{\rm diff}-H_{\Lambda{\rm CDM}}}
     {H_{\Lambda{\rm CDM}}}.
\end{equation}
The two expansion histories are nearly indistinguishable over the entire redshift range $0\le z\le3$, illustrating that the diffusion sector preserves the overall background evolution required by current observations.

The lower panel reveals that the diffusion-induced modification remains at the sub-percent level throughout the considered redshift interval. The largest departure occurs around intermediate redshifts, $z\simeq0.5$, where the relative difference reaches approximately one percent. At both lower and higher redshifts the deviation rapidly decreases, approaching zero as the diffusion contribution becomes either dynamically sud-dominant or negligible relative to the total energy budget. This behavior demonstrates that the diffusion sector primarily affects the recent expansion history while leaving the high-redshift evolution essentially unchanged.

An important feature of the residual panel is the sign of $\Delta H/H_{\Lambda{\rm CDM}}$, which directly indicates whether diffusion predicts a faster or slower expansion rate than the corresponding $\Lambda$CDM solution. Positive values correspond to a larger expansion rate, whereas negative values indicate a slower expansion. The residual curve changes sign across the explored redshift range, implying that \textit{the diffusion sector does not simply introduce a uniform shift in $H(z)$ but rather produces a redshift-dependent modification of the expansion history}. This compensating behavior explains why the overall deviation remains extremely small despite the presence of a non-negligible diffusion component.

\section{Conclusions}
\label{sec:conclusions}
In this paper, we have provided clear insights that current observational data indicate that the diffusion sector remains observationally viable and consistently improves the quality of the fit when additional datasets are included. 
Our results are very robust under changes in the SNIa sample used in the analysis. However, the statistical significance of this improvement must be weighed against the increased model complexity. While the DESY5 compilation yields a moderate preference for diffusion according to the Akaike Information Criterion, the more recent DES-Dovekie and PantheonPlus samples do not provide sufficient improvement in $\chi^2$ to compensate for the three additional model parameters. Consequently, the Bayesian Information Criterion consistently favors $\Lambda$CDM over the general diffusion scenario considered in this work.

It is worth emphasizing, however, that the present analysis explores the most general realization of the diffusion framework. Simpler and more predictive subclasses, such as constant-slope diffusion ($\Gamma_Q=1$), corresponding to the power-law form $Q\propto a^{-\lambda}$, reduce the number of free parameters by fixing the diffusion curvature and slope evolution. Such reduced models may provide a more favorable balance between goodness of fit and model complexity, potentially leading to smaller AIC and BIC values. Furthermore, a pure diffusion scenario in which the cosmological constant contribution is absent ($\Omega_\Lambda=0$) represents an interesting limiting case that deserves further investigation. Future analyses, together with forthcoming cosmological data providing either a larger improvement in $\chi^2$ or tighter constraints on the diffusion sector, will be essential to determine whether diffusion cosmologies can achieve a statistically significant preference over $\Lambda$CDM.

Interestingly, statistical analysis point out that the preferred present-day diffusion density remains consistently around
$\Omega_{Q0}\simeq 0.2$, while the fractional contribution of
the diffusion component to the dark-energy sector is constrained to $f_Q\simeq 0.3$--$0.4$, with a marginalized posterior distribution clearly indicating a mean value away from zero. The constraints summarized in Table~\ref{tab:diffusion_constraints}
show a remarkable stability of the diffusion sector across all dataset combinations considered.

In all dataset combinations considered, the posterior distribution of $\Gamma_{0}$ is broad and significantly non-Gaussian, with median values consistently below unity. The inferred values favor $\Gamma_{0}\approx 0.45$, indicating that, within the generalized diffusion framework, the data prefer a running diffusion slope over the constant-slope power-law case ($\Gamma_{Q}=1$). However, the preference is not sufficiently strong to exclude the power-law limit at high confidence, and should therefore be interpreted as a mild preference for a more general diffusion evolution rather than as evidence against power-law diffusion.

The fact that the diffusion model improves the overall fit to the combined DESI DR2 + PantheonPlus + CC dataset while modifying the Hubble rate by less than one percent is a highly nontrivial result. It shows that the diffusion sector can leave a measurable imprint on the recent cosmological dynamics without spoiling the excellent agreement with current cosmological probes. In this sense, diffusion acts as a subtle deformation of the dark-energy sector, capable of improving the statistical fit while remaining fully compatible with existing observations of the expansion history.

These results motivate extending the analysis beyond the background cosmology by incorporating probes of structure formation, such as redshift-space distortions, weak lensing, and galaxy clustering. While low-redshift background observations provide the primary constraints on the late-time evolution of the diffusion sector, perturbative observables offer a complementary avenue to test the impact of the dark-sector energy exchange on the growth of cosmic structures. Such analyses will be essential to determine whether diffusion represents a genuine cosmological component with distinctive signatures beyond the background level or merely an effective description compatible with current observations.

\section*{Acknowledgments}
Financial support from the Chilean National Agency for Research and Development (ANID) through Fondecyt Grant No. 1250969 is gratefully acknowledged.

\bibliography{biblio.bib}

\begin{thebibliography}{47}%
\makeatletter
\providecommand \@ifxundefined [1]{%
 \@ifx{#1\undefined}
}%
\providecommand \@ifnum [1]{%
 \ifnum #1\expandafter \@firstoftwo
 \else \expandafter \@secondoftwo
 \fi
}%
\providecommand \@ifx [1]{%
 \ifx #1\expandafter \@firstoftwo
 \else \expandafter \@secondoftwo
 \fi
}%
\providecommand \natexlab [1]{#1}%
\providecommand \enquote  [1]{``#1''}%
\providecommand \bibnamefont  [1]{#1}%
\providecommand \bibfnamefont [1]{#1}%
\providecommand \citenamefont [1]{#1}%
\providecommand \href@noop [0]{\@secondoftwo}%
\providecommand \href [0]{\begingroup \@sanitize@url \@href}%
\providecommand \@href[1]{\@@startlink{#1}\@@href}%
\providecommand \@@href[1]{\endgroup#1\@@endlink}%
\providecommand \@sanitize@url [0]{\catcode `\\12\catcode `\$12\catcode `\&12\catcode `\#12\catcode `\^12\catcode `\_12\catcode `\%12\relax}%
\providecommand \@@startlink[1]{}%
\providecommand \@@endlink[0]{}%
\providecommand \url  [0]{\begingroup\@sanitize@url \@url }%
\providecommand \@url [1]{\endgroup\@href {#1}{\urlprefix }}%
\providecommand \urlprefix  [0]{URL }%
\providecommand \Eprint [0]{\href }%
\providecommand \doibase [0]{http://dx.doi.org/}%
\providecommand \selectlanguage [0]{\@gobble}%
\providecommand \bibinfo  [0]{\@secondoftwo}%
\providecommand \bibfield  [0]{\@secondoftwo}%
\providecommand \translation [1]{[#1]}%
\providecommand \BibitemOpen [0]{}%
\providecommand \bibitemStop [0]{}%
\providecommand \bibitemNoStop [0]{.\EOS\space}%
\providecommand \EOS [0]{\spacefactor3000\relax}%
\providecommand \BibitemShut  [1]{\csname bibitem#1\endcsname}%
\let\auto@bib@innerbib\@empty
\bibitem [{\citenamefont {Blanchard}\ \emph {et~al.}(2018)\citenamefont {Blanchard}, \citenamefont {Sakr},\ and\ \citenamefont {Ilic}}]{Blanchard:2018klb}%
  \BibitemOpen
  \bibfield  {author} {\bibinfo {author} {\bibfnamefont {A.}~\bibnamefont {Blanchard}}, \bibinfo {author} {\bibfnamefont {Z.}~\bibnamefont {Sakr}}, \ and\ \bibinfo {author} {\bibfnamefont {S.}~\bibnamefont {Ilic}},\ }in\ \href@noop {} {\emph {\bibinfo {booktitle} {{53rd Rencontres de Moriond on Cosmology}}}}\ (\bibinfo {year} {2018})\ pp.\ \bibinfo {pages} {39--42},\ \Eprint {http://arxiv.org/abs/1805.06976} {arXiv:1805.06976 [astro-ph.CO]} \BibitemShut {NoStop}%
\bibitem [{\citenamefont {Efstathiou}(2025)}]{Efstathiou:2024dvn}%
  \BibitemOpen
  \bibfield  {author} {\bibinfo {author} {\bibfnamefont {G.}~\bibnamefont {Efstathiou}},\ }\href {\doibase 10.1098/rsta.2024.0022} {\bibfield  {journal} {\bibinfo  {journal} {Phil. Trans. Roy. Soc. Lond. A}\ }\textbf {\bibinfo {volume} {383}},\ \bibinfo {pages} {20240022} (\bibinfo {year} {2025})},\ \Eprint {http://arxiv.org/abs/2406.12106} {arXiv:2406.12106 [astro-ph.CO]} \BibitemShut {NoStop}%
\bibitem [{\citenamefont {Riess}\ \emph {et~al.}(2022)\citenamefont {Riess} \emph {et~al.}}]{Riess:2021jrx}%
  \BibitemOpen
  \bibfield  {author} {\bibinfo {author} {\bibfnamefont {A.~G.}\ \bibnamefont {Riess}} \emph {et~al.},\ }\href {\doibase 10.3847/2041-8213/ac5c5b} {\bibfield  {journal} {\bibinfo  {journal} {Astrophys. J. Lett.}\ }\textbf {\bibinfo {volume} {934}},\ \bibinfo {pages} {L7} (\bibinfo {year} {2022})},\ \Eprint {http://arxiv.org/abs/2112.04510} {arXiv:2112.04510 [astro-ph.CO]} \BibitemShut {NoStop}%
\bibitem [{\citenamefont {Knox}\ and\ \citenamefont {Millea}(2020)}]{Knox:2019rjx}%
  \BibitemOpen
  \bibfield  {author} {\bibinfo {author} {\bibfnamefont {L.}~\bibnamefont {Knox}}\ and\ \bibinfo {author} {\bibfnamefont {M.}~\bibnamefont {Millea}},\ }\href {\doibase 10.1103/PhysRevD.101.043533} {\bibfield  {journal} {\bibinfo  {journal} {Phys. Rev. D}\ }\textbf {\bibinfo {volume} {101}},\ \bibinfo {pages} {043533} (\bibinfo {year} {2020})},\ \Eprint {http://arxiv.org/abs/1908.03663} {arXiv:1908.03663 [astro-ph.CO]} \BibitemShut {NoStop}%
\bibitem [{\citenamefont {Casertano}\ \emph {et~al.}(2026)\citenamefont {Casertano} \emph {et~al.}}]{H0DN:2025lyy}%
  \BibitemOpen
  \bibfield  {author} {\bibinfo {author} {\bibfnamefont {S.}~\bibnamefont {Casertano}} \emph {et~al.} (\bibinfo {collaboration} {H0DN}),\ }\href {\doibase 10.1051/0004-6361/202557993} {\bibfield  {journal} {\bibinfo  {journal} {Astron. Astrophys.}\ }\textbf {\bibinfo {volume} {708}},\ \bibinfo {pages} {A166} (\bibinfo {year} {2026})},\ \Eprint {http://arxiv.org/abs/2510.23823} {arXiv:2510.23823 [astro-ph.CO]} \BibitemShut {NoStop}%
\bibitem [{\citenamefont {Verde}\ \emph {et~al.}(2024)\citenamefont {Verde}, \citenamefont {Sch{\"o}neberg},\ and\ \citenamefont {Gil-Mar{\'\i}n}}]{verde2024tale}%
  \BibitemOpen
  \bibfield  {author} {\bibinfo {author} {\bibfnamefont {L.}~\bibnamefont {Verde}}, \bibinfo {author} {\bibfnamefont {N.}~\bibnamefont {Sch{\"o}neberg}}, \ and\ \bibinfo {author} {\bibfnamefont {H.}~\bibnamefont {Gil-Mar{\'\i}n}},\ }\href@noop {} {\bibfield  {journal} {\bibinfo  {journal} {Annual Review of Astronomy and Astrophysics}\ }\textbf {\bibinfo {volume} {62}},\ \bibinfo {pages} {287} (\bibinfo {year} {2024})}\BibitemShut {NoStop}%
\bibitem [{\citenamefont {Kamionkowski}\ and\ \citenamefont {Riess}(2023)}]{kamionkowski2023hubble}%
  \BibitemOpen
  \bibfield  {author} {\bibinfo {author} {\bibfnamefont {M.}~\bibnamefont {Kamionkowski}}\ and\ \bibinfo {author} {\bibfnamefont {A.~G.}\ \bibnamefont {Riess}},\ }\href@noop {} {\bibfield  {journal} {\bibinfo  {journal} {Annual Review of Nuclear and Particle Science}\ }\textbf {\bibinfo {volume} {73}},\ \bibinfo {pages} {153} (\bibinfo {year} {2023})}\BibitemShut {NoStop}%
\bibitem [{\citenamefont {Macaulay}\ \emph {et~al.}(2013)\citenamefont {Macaulay}, \citenamefont {Wehus},\ and\ \citenamefont {Eriksen}}]{Macaulay:2013swa}%
  \BibitemOpen
  \bibfield  {author} {\bibinfo {author} {\bibfnamefont {E.}~\bibnamefont {Macaulay}}, \bibinfo {author} {\bibfnamefont {I.~K.}\ \bibnamefont {Wehus}}, \ and\ \bibinfo {author} {\bibfnamefont {H.~K.}\ \bibnamefont {Eriksen}},\ }\href {\doibase 10.1103/PhysRevLett.111.161301} {\bibfield  {journal} {\bibinfo  {journal} {Phys. Rev. Lett.}\ }\textbf {\bibinfo {volume} {111}},\ \bibinfo {pages} {161301} (\bibinfo {year} {2013})},\ \Eprint {http://arxiv.org/abs/1303.6583} {arXiv:1303.6583 [astro-ph.CO]} \BibitemShut {NoStop}%
\bibitem [{\citenamefont {Battye}\ \emph {et~al.}(2015)\citenamefont {Battye}, \citenamefont {Charnock},\ and\ \citenamefont {Moss}}]{Battye:2014qga}%
  \BibitemOpen
  \bibfield  {author} {\bibinfo {author} {\bibfnamefont {R.~A.}\ \bibnamefont {Battye}}, \bibinfo {author} {\bibfnamefont {T.}~\bibnamefont {Charnock}}, \ and\ \bibinfo {author} {\bibfnamefont {A.}~\bibnamefont {Moss}},\ }\href {\doibase 10.1103/PhysRevD.91.103508} {\bibfield  {journal} {\bibinfo  {journal} {Phys. Rev. D}\ }\textbf {\bibinfo {volume} {91}},\ \bibinfo {pages} {103508} (\bibinfo {year} {2015})},\ \Eprint {http://arxiv.org/abs/1409.2769} {arXiv:1409.2769 [astro-ph.CO]} \BibitemShut {NoStop}%
\bibitem [{\citenamefont {Alam}\ \emph {et~al.}(2017)\citenamefont {Alam} \emph {et~al.}}]{Alam:2016hwk}%
  \BibitemOpen
  \bibfield  {author} {\bibinfo {author} {\bibfnamefont {S.}~\bibnamefont {Alam}} \emph {et~al.} (\bibinfo {collaboration} {BOSS}),\ }\href {\doibase 10.1093/mnras/stx721} {\bibfield  {journal} {\bibinfo  {journal} {Mon. Not. Roy. Astron. Soc.}\ }\textbf {\bibinfo {volume} {470}},\ \bibinfo {pages} {2617} (\bibinfo {year} {2017})},\ \Eprint {http://arxiv.org/abs/1607.03155} {arXiv:1607.03155 [astro-ph.CO]} \BibitemShut {NoStop}%
\bibitem [{\citenamefont {Abbott}\ \emph {et~al.}(2018)\citenamefont {Abbott} \emph {et~al.}}]{Abbott:2017wau}%
  \BibitemOpen
  \bibfield  {author} {\bibinfo {author} {\bibfnamefont {T.~M.~C.}\ \bibnamefont {Abbott}} \emph {et~al.} (\bibinfo {collaboration} {DES}),\ }\href {\doibase 10.1103/PhysRevD.98.043526} {\bibfield  {journal} {\bibinfo  {journal} {Phys. Rev. D}\ }\textbf {\bibinfo {volume} {98}},\ \bibinfo {pages} {043526} (\bibinfo {year} {2018})},\ \Eprint {http://arxiv.org/abs/1708.01530} {arXiv:1708.01530 [astro-ph.CO]} \BibitemShut {NoStop}%
\bibitem [{\citenamefont {Abdul~Karim}\ \emph {et~al.}(2025)\citenamefont {Abdul~Karim} \emph {et~al.}}]{DESI:2025zgx}%
  \BibitemOpen
  \bibfield  {author} {\bibinfo {author} {\bibfnamefont {M.}~\bibnamefont {Abdul~Karim}} \emph {et~al.} (\bibinfo {collaboration} {DESI}),\ }\href {\doibase 10.1103/tr6y-kpc6} {\bibfield  {journal} {\bibinfo  {journal} {Phys. Rev. D}\ }\textbf {\bibinfo {volume} {112}},\ \bibinfo {pages} {083515} (\bibinfo {year} {2025})},\ \Eprint {http://arxiv.org/abs/2503.14738} {arXiv:2503.14738 [astro-ph.CO]} \BibitemShut {NoStop}%
\bibitem [{\citenamefont {Gu}\ \emph {et~al.}(2025)\citenamefont {Gu} \emph {et~al.}}]{DESI:2025wyn}%
  \BibitemOpen
  \bibfield  {author} {\bibinfo {author} {\bibfnamefont {G.}~\bibnamefont {Gu}} \emph {et~al.} (\bibinfo {collaboration} {DESI}),\ }\href {\doibase 10.1038/s41550-025-02669-6} {\bibfield  {journal} {\bibinfo  {journal} {Nature Astron.}\ }\textbf {\bibinfo {volume} {9}},\ \bibinfo {pages} {1879} (\bibinfo {year} {2025})},\ \bibinfo {note} {[Erratum: Nature Astron. 9, 1898--1898 (2025)]},\ \Eprint {http://arxiv.org/abs/2504.06118} {arXiv:2504.06118 [astro-ph.CO]} \BibitemShut {NoStop}%
\bibitem [{\citenamefont {Capozziello}\ \emph {et~al.}(2025)\citenamefont {Capozziello}, \citenamefont {Chaudhary}, \citenamefont {Harko},\ and\ \citenamefont {Mustafa}}]{capozziello2025dark}%
  \BibitemOpen
  \bibfield  {author} {\bibinfo {author} {\bibfnamefont {S.}~\bibnamefont {Capozziello}}, \bibinfo {author} {\bibfnamefont {H.}~\bibnamefont {Chaudhary}}, \bibinfo {author} {\bibfnamefont {T.}~\bibnamefont {Harko}}, \ and\ \bibinfo {author} {\bibfnamefont {G.}~\bibnamefont {Mustafa}},\ }\href@noop {} {\bibfield  {journal} {\bibinfo  {journal} {Physics of the Dark Universe}\ ,\ \bibinfo {pages} {102196}} (\bibinfo {year} {2025})}\BibitemShut {NoStop}%
\bibitem [{\citenamefont {Ong}\ \emph {et~al.}(2025)\citenamefont {Ong}, \citenamefont {Yallup},\ and\ \citenamefont {Handley}}]{Ong:2025utx}%
  \BibitemOpen
  \bibfield  {author} {\bibinfo {author} {\bibfnamefont {D.~D.~Y.}\ \bibnamefont {Ong}}, \bibinfo {author} {\bibfnamefont {D.}~\bibnamefont {Yallup}}, \ and\ \bibinfo {author} {\bibfnamefont {W.}~\bibnamefont {Handley}},\ }\href@noop {} {\  (\bibinfo {year} {2025})},\ \Eprint {http://arxiv.org/abs/2511.10631} {arXiv:2511.10631 [astro-ph.CO]} \BibitemShut {NoStop}%
\bibitem [{\citenamefont {Dinda}\ \emph {et~al.}(2026)\citenamefont {Dinda}, \citenamefont {Maartens},\ and\ \citenamefont {Saito}}]{Dinda:2026ktu}%
  \BibitemOpen
  \bibfield  {author} {\bibinfo {author} {\bibfnamefont {B.~R.}\ \bibnamefont {Dinda}}, \bibinfo {author} {\bibfnamefont {R.}~\bibnamefont {Maartens}}, \ and\ \bibinfo {author} {\bibfnamefont {S.}~\bibnamefont {Saito}},\ }\href@noop {} {\  (\bibinfo {year} {2026})},\ \Eprint {http://arxiv.org/abs/2605.13546} {arXiv:2605.13546 [astro-ph.CO]} \BibitemShut {NoStop}%
\bibitem [{\citenamefont {Cai}\ and\ \citenamefont {Wang}(2026)}]{cai2026hubble}%
  \BibitemOpen
  \bibfield  {author} {\bibinfo {author} {\bibfnamefont {R.-G.}\ \bibnamefont {Cai}}\ and\ \bibinfo {author} {\bibfnamefont {S.-J.}\ \bibnamefont {Wang}},\ }\href@noop {} {\bibfield  {journal} {\bibinfo  {journal} {Research in Astronomy and Astrophysics}\ }\textbf {\bibinfo {volume} {26}},\ \bibinfo {pages} {084011} (\bibinfo {year} {2026})}\BibitemShut {NoStop}%
\bibitem [{\citenamefont {Jia}\ \emph {et~al.}(2026)\citenamefont {Jia}, \citenamefont {Dai}, \citenamefont {Yang},\ and\ \citenamefont {Wang}}]{jia2026review}%
  \BibitemOpen
  \bibfield  {author} {\bibinfo {author} {\bibfnamefont {X.-D.}\ \bibnamefont {Jia}}, \bibinfo {author} {\bibfnamefont {X.-Y.}\ \bibnamefont {Dai}}, \bibinfo {author} {\bibfnamefont {Y.-P.}\ \bibnamefont {Yang}}, \ and\ \bibinfo {author} {\bibfnamefont {F.-Y.}\ \bibnamefont {Wang}},\ }\href@noop {} {\bibfield  {journal} {\bibinfo  {journal} {Galaxies}\ }\textbf {\bibinfo {volume} {14}},\ \bibinfo {pages} {55} (\bibinfo {year} {2026})}\BibitemShut {NoStop}%
\bibitem [{\citenamefont {Landau}\ \emph {et~al.}(2023)\citenamefont {Landau}, \citenamefont {Benetti}, \citenamefont {Perez},\ and\ \citenamefont {Sudarsky}}]{landau2023cosmological}%
  \BibitemOpen
  \bibfield  {author} {\bibinfo {author} {\bibfnamefont {S.~J.}\ \bibnamefont {Landau}}, \bibinfo {author} {\bibfnamefont {M.}~\bibnamefont {Benetti}}, \bibinfo {author} {\bibfnamefont {A.}~\bibnamefont {Perez}}, \ and\ \bibinfo {author} {\bibfnamefont {D.}~\bibnamefont {Sudarsky}},\ }\href@noop {} {\bibfield  {journal} {\bibinfo  {journal} {Physical Review D}\ }\textbf {\bibinfo {volume} {108}},\ \bibinfo {pages} {043524} (\bibinfo {year} {2023})}\BibitemShut {NoStop}%
\bibitem [{\citenamefont {Perez}\ \emph {et~al.}(2021)\citenamefont {Perez}, \citenamefont {Sudarsky},\ and\ \citenamefont {Wilson-Ewing}}]{perez2021resolving}%
  \BibitemOpen
  \bibfield  {author} {\bibinfo {author} {\bibfnamefont {A.}~\bibnamefont {Perez}}, \bibinfo {author} {\bibfnamefont {D.}~\bibnamefont {Sudarsky}}, \ and\ \bibinfo {author} {\bibfnamefont {E.}~\bibnamefont {Wilson-Ewing}},\ }\href@noop {} {\bibfield  {journal} {\bibinfo  {journal} {General Relativity and Gravitation}\ }\textbf {\bibinfo {volume} {53}},\ \bibinfo {pages} {7} (\bibinfo {year} {2021})}\BibitemShut {NoStop}%
\bibitem [{\citenamefont {Linares Cede\~no}\ and\ \citenamefont {Nucamendi}(2021)}]{LinaresCedeno:2020uxx}%
  \BibitemOpen
  \bibfield  {author} {\bibinfo {author} {\bibfnamefont {F.~X.}\ \bibnamefont {Linares Cede\~no}}\ and\ \bibinfo {author} {\bibfnamefont {U.}~\bibnamefont {Nucamendi}},\ }\href {\doibase 10.1016/j.dark.2021.100807} {\bibfield  {journal} {\bibinfo  {journal} {Phys. Dark Univ.}\ }\textbf {\bibinfo {volume} {32}},\ \bibinfo {pages} {100807} (\bibinfo {year} {2021})}\BibitemShut {NoStop}%
\bibitem [{\citenamefont {Singh}\ and\ \citenamefont {Kashyap}(2023)}]{singh2023unimodular}%
  \BibitemOpen
  \bibfield  {author} {\bibinfo {author} {\bibfnamefont {N.~K.}\ \bibnamefont {Singh}}\ and\ \bibinfo {author} {\bibfnamefont {G.}~\bibnamefont {Kashyap}},\ }\href@noop {} {\bibfield  {journal} {\bibinfo  {journal} {Universe}\ }\textbf {\bibinfo {volume} {9}},\ \bibinfo {pages} {469} (\bibinfo {year} {2023})}\BibitemShut {NoStop}%
\bibitem [{\citenamefont {Pearle}(1976)}]{Pearle:1976ka}%
  \BibitemOpen
  \bibfield  {author} {\bibinfo {author} {\bibfnamefont {P.~M.}\ \bibnamefont {Pearle}},\ }\href {\doibase 10.1103/PhysRevD.13.857} {\bibfield  {journal} {\bibinfo  {journal} {Phys. Rev.}\ }\textbf {\bibinfo {volume} {D13}},\ \bibinfo {pages} {857} (\bibinfo {year} {1976})}\BibitemShut {NoStop}%
\bibitem [{\citenamefont {Ghirardi}\ \emph {et~al.}(1986)\citenamefont {Ghirardi}, \citenamefont {Rimini},\ and\ \citenamefont {Weber}}]{Ghirardi:1985mt}%
  \BibitemOpen
  \bibfield  {author} {\bibinfo {author} {\bibfnamefont {G.~C.}\ \bibnamefont {Ghirardi}}, \bibinfo {author} {\bibfnamefont {A.}~\bibnamefont {Rimini}}, \ and\ \bibinfo {author} {\bibfnamefont {T.}~\bibnamefont {Weber}},\ }\href {\doibase 10.1103/PhysRevD.34.470} {\bibfield  {journal} {\bibinfo  {journal} {Phys. Rev.}\ }\textbf {\bibinfo {volume} {D34}},\ \bibinfo {pages} {470} (\bibinfo {year} {1986})}\BibitemShut {NoStop}%
\bibitem [{\citenamefont {Pearle}(1989)}]{Pearle:1988uh}%
  \BibitemOpen
  \bibfield  {author} {\bibinfo {author} {\bibfnamefont {P.~M.}\ \bibnamefont {Pearle}},\ }\href {\doibase 10.1103/PhysRevA.39.2277} {\bibfield  {journal} {\bibinfo  {journal} {Phys. Rev.}\ }\textbf {\bibinfo {volume} {A39}},\ \bibinfo {pages} {2277} (\bibinfo {year} {1989})}\BibitemShut {NoStop}%
\bibitem [{\citenamefont {Ghirardi}\ \emph {et~al.}(1990)\citenamefont {Ghirardi}, \citenamefont {Pearle},\ and\ \citenamefont {Rimini}}]{Ghirardi:1989cn}%
  \BibitemOpen
  \bibfield  {author} {\bibinfo {author} {\bibfnamefont {G.~C.}\ \bibnamefont {Ghirardi}}, \bibinfo {author} {\bibfnamefont {P.~M.}\ \bibnamefont {Pearle}}, \ and\ \bibinfo {author} {\bibfnamefont {A.}~\bibnamefont {Rimini}},\ }\href {\doibase 10.1103/PhysRevA.42.78} {\bibfield  {journal} {\bibinfo  {journal} {Phys. Rev.}\ }\textbf {\bibinfo {volume} {A42}},\ \bibinfo {pages} {78} (\bibinfo {year} {1990})}\BibitemShut {NoStop}%
\bibitem [{\citenamefont {Garc{\'\i}a-Aspeitia}\ \emph {et~al.}(2019)\citenamefont {Garc{\'\i}a-Aspeitia}, \citenamefont {Mart{\'\i}nez-Robles}, \citenamefont {Hern{\'a}ndez-Almada}, \citenamefont {Maga{\~n}a},\ and\ \citenamefont {Motta}}]{garcia2019cosmic}%
  \BibitemOpen
  \bibfield  {author} {\bibinfo {author} {\bibfnamefont {M.~A.}\ \bibnamefont {Garc{\'\i}a-Aspeitia}}, \bibinfo {author} {\bibfnamefont {C.}~\bibnamefont {Mart{\'\i}nez-Robles}}, \bibinfo {author} {\bibfnamefont {A.}~\bibnamefont {Hern{\'a}ndez-Almada}}, \bibinfo {author} {\bibfnamefont {J.}~\bibnamefont {Maga{\~n}a}}, \ and\ \bibinfo {author} {\bibfnamefont {V.}~\bibnamefont {Motta}},\ }\href@noop {} {\bibfield  {journal} {\bibinfo  {journal} {Physical Review D}\ }\textbf {\bibinfo {volume} {99}},\ \bibinfo {pages} {123525} (\bibinfo {year} {2019})}\BibitemShut {NoStop}%
\bibitem [{\citenamefont {Garc{\'\i}a-Aspeitia}\ \emph {et~al.}(2021)\citenamefont {Garc{\'\i}a-Aspeitia}, \citenamefont {Hern{\'a}ndez-Almada}, \citenamefont {Maga{\~n}a},\ and\ \citenamefont {Motta}}]{garcia2021universe}%
  \BibitemOpen
  \bibfield  {author} {\bibinfo {author} {\bibfnamefont {M.~A.}\ \bibnamefont {Garc{\'\i}a-Aspeitia}}, \bibinfo {author} {\bibfnamefont {A.}~\bibnamefont {Hern{\'a}ndez-Almada}}, \bibinfo {author} {\bibfnamefont {J.}~\bibnamefont {Maga{\~n}a}}, \ and\ \bibinfo {author} {\bibfnamefont {V.}~\bibnamefont {Motta}},\ }\href@noop {} {\bibfield  {journal} {\bibinfo  {journal} {Physics of the Dark Universe}\ }\textbf {\bibinfo {volume} {32}},\ \bibinfo {pages} {100840} (\bibinfo {year} {2021})}\BibitemShut {NoStop}%
\bibitem [{\citenamefont {Zafari}\ \emph {et~al.}(2026)\citenamefont {Zafari}, \citenamefont {Namdar}, \citenamefont {Escamilla},\ and\ \citenamefont {Di~Valentino}}]{zafari2026testing}%
  \BibitemOpen
  \bibfield  {author} {\bibinfo {author} {\bibfnamefont {A.}~\bibnamefont {Zafari}}, \bibinfo {author} {\bibfnamefont {M.-H.}\ \bibnamefont {Namdar}}, \bibinfo {author} {\bibfnamefont {L.~A.}\ \bibnamefont {Escamilla}}, \ and\ \bibinfo {author} {\bibfnamefont {E.}~\bibnamefont {Di~Valentino}},\ }\href@noop {} {\bibfield  {journal} {\bibinfo  {journal} {arXiv preprint arXiv:2608.15950}\ } (\bibinfo {year} {2026})}\BibitemShut {NoStop}%
\bibitem [{\citenamefont {G{\'o}mez}\ \emph {et~al.}(2026)\citenamefont {G{\'o}mez}, \citenamefont {Palma},\ and\ \citenamefont {Cruz}}]{Gomez:2026eyp}%
  \BibitemOpen
  \bibfield  {author} {\bibinfo {author} {\bibfnamefont {G.}~\bibnamefont {G{\'o}mez}}, \bibinfo {author} {\bibfnamefont {G.}~\bibnamefont {Palma}}, \ and\ \bibinfo {author} {\bibfnamefont {N.}~\bibnamefont {Cruz}},\ }\href@noop {} {\  (\bibinfo {year} {2026})},\ \Eprint {http://arxiv.org/abs/2609.00201} {arXiv:2609.00201 [gr-qc]} \BibitemShut {NoStop}%
\bibitem [{\citenamefont {Corral}\ \emph {et~al.}(2020)\citenamefont {Corral}, \citenamefont {Cruz},\ and\ \citenamefont {Gonz\'alez}}]{Corral:2020lxt}%
  \BibitemOpen
  \bibfield  {author} {\bibinfo {author} {\bibfnamefont {C.}~\bibnamefont {Corral}}, \bibinfo {author} {\bibfnamefont {N.}~\bibnamefont {Cruz}}, \ and\ \bibinfo {author} {\bibfnamefont {E.}~\bibnamefont {Gonz\'alez}},\ }\href {\doibase 10.1103/PhysRevD.102.023508} {\bibfield  {journal} {\bibinfo  {journal} {Phys. Rev. D}\ }\textbf {\bibinfo {volume} {102}},\ \bibinfo {pages} {023508} (\bibinfo {year} {2020})}\BibitemShut {NoStop}%
\bibitem [{\citenamefont {Cruz}\ \emph {et~al.}(2026)\citenamefont {Cruz}, \citenamefont {Lepe}, \citenamefont {Palma},\ and\ \citenamefont {Cruz}}]{cruz2026thermodynamicconstraintsfuturesingularities}%
  \BibitemOpen
  \bibfield  {author} {\bibinfo {author} {\bibfnamefont {N.}~\bibnamefont {Cruz}}, \bibinfo {author} {\bibfnamefont {S.}~\bibnamefont {Lepe}}, \bibinfo {author} {\bibfnamefont {G.}~\bibnamefont {Palma}}, \ and\ \bibinfo {author} {\bibfnamefont {M.}~\bibnamefont {Cruz}},\ }\href {https://arxiv.org/abs/2603.22749} {\enquote {\bibinfo {title} {Thermodynamic constraints and future singularities in unimodular gravity driven by phantom and non-phantom fluids},}\ } (\bibinfo {year} {2026}),\ \Eprint {http://arxiv.org/abs/2603.22749} {arXiv:2603.22749 [gr-qc]} \BibitemShut {NoStop}%
\bibitem [{\citenamefont {Einstein}(1919)}]{Einstein:1919gv}%
  \BibitemOpen
  \bibfield  {author} {\bibinfo {author} {\bibfnamefont {A.}~\bibnamefont {Einstein}},\ }\href@noop {} {\bibfield  {journal} {\bibinfo  {journal} {Sitzungsber. Preuss. Akad. Wiss. Berlin (Math. Phys. )}\ }\textbf {\bibinfo {volume} {1919}},\ \bibinfo {pages} {349} (\bibinfo {year} {1919})}\BibitemShut {NoStop}%
\bibitem [{\citenamefont {Torrado}\ and\ \citenamefont {Lewis}(2021)}]{Torrado:2020dgo}%
  \BibitemOpen
  \bibfield  {author} {\bibinfo {author} {\bibfnamefont {J.}~\bibnamefont {Torrado}}\ and\ \bibinfo {author} {\bibfnamefont {A.}~\bibnamefont {Lewis}},\ }\href {\doibase 10.1088/1475-7516/2021/05/057} {\bibfield  {journal} {\bibinfo  {journal} {JCAP}\ }\textbf {\bibinfo {volume} {05}},\ \bibinfo {pages} {057} (\bibinfo {year} {2021})},\ \Eprint {http://arxiv.org/abs/2005.05290} {arXiv:2005.05290 [astro-ph.IM]} \BibitemShut {NoStop}%
\bibitem [{\citenamefont {Gelman}\ and\ \citenamefont {Rubin}(1992)}]{Gelman:1992zz}%
  \BibitemOpen
  \bibfield  {author} {\bibinfo {author} {\bibfnamefont {A.}~\bibnamefont {Gelman}}\ and\ \bibinfo {author} {\bibfnamefont {D.~B.}\ \bibnamefont {Rubin}},\ }\href {\doibase 10.1214/ss/1177011136} {\bibfield  {journal} {\bibinfo  {journal} {Statist. Sci.}\ }\textbf {\bibinfo {volume} {7}},\ \bibinfo {pages} {457} (\bibinfo {year} {1992})}\BibitemShut {NoStop}%
\bibitem [{\citenamefont {Lewis}(2025)}]{Lewis:2019xzd}%
  \BibitemOpen
  \bibfield  {author} {\bibinfo {author} {\bibfnamefont {A.}~\bibnamefont {Lewis}},\ }\href {\doibase 10.1088/1475-7516/2025/08/025} {\bibfield  {journal} {\bibinfo  {journal} {JCAP}\ }\textbf {\bibinfo {volume} {08}},\ \bibinfo {pages} {025} (\bibinfo {year} {2025})},\ \Eprint {http://arxiv.org/abs/1910.13970} {arXiv:1910.13970 [astro-ph.IM]} \BibitemShut {NoStop}%
\bibitem [{\citenamefont {Roy}\ and\ \citenamefont {Sahoo}(2026)}]{Roy:2026icy}%
  \BibitemOpen
  \bibfield  {author} {\bibinfo {author} {\bibfnamefont {N.}~\bibnamefont {Roy}}\ and\ \bibinfo {author} {\bibfnamefont {P.}~\bibnamefont {Sahoo}},\ }\href@noop {} {\  (\bibinfo {year} {2026})},\ \Eprint {http://arxiv.org/abs/2603.14740} {arXiv:2603.14740 [astro-ph.CO]} \BibitemShut {NoStop}%
\bibitem [{\citenamefont {Hogg}(1999)}]{Hogg:1999ad}%
  \BibitemOpen
  \bibfield  {author} {\bibinfo {author} {\bibfnamefont {D.~W.}\ \bibnamefont {Hogg}},\ }\href@noop {} {\  (\bibinfo {year} {1999})},\ \Eprint {http://arxiv.org/abs/astro-ph/9905116} {arXiv:astro-ph/9905116} \BibitemShut {NoStop}%
\bibitem [{\citenamefont {Weinberg}\ \emph {et~al.}(2013)\citenamefont {Weinberg}, \citenamefont {Mortonson}, \citenamefont {Eisenstein}, \citenamefont {Hirata}, \citenamefont {Riess},\ and\ \citenamefont {Rozo}}]{Weinberg:2013agg}%
  \BibitemOpen
  \bibfield  {author} {\bibinfo {author} {\bibfnamefont {D.~H.}\ \bibnamefont {Weinberg}}, \bibinfo {author} {\bibfnamefont {M.~J.}\ \bibnamefont {Mortonson}}, \bibinfo {author} {\bibfnamefont {D.~J.}\ \bibnamefont {Eisenstein}}, \bibinfo {author} {\bibfnamefont {C.}~\bibnamefont {Hirata}}, \bibinfo {author} {\bibfnamefont {A.~G.}\ \bibnamefont {Riess}}, \ and\ \bibinfo {author} {\bibfnamefont {E.}~\bibnamefont {Rozo}},\ }\href {\doibase 10.1016/j.physrep.2013.05.001} {\bibfield  {journal} {\bibinfo  {journal} {Phys. Rept.}\ }\textbf {\bibinfo {volume} {530}},\ \bibinfo {pages} {87} (\bibinfo {year} {2013})},\ \Eprint {http://arxiv.org/abs/1201.2434} {arXiv:1201.2434 [astro-ph.CO]} \BibitemShut {NoStop}%
\bibitem [{\citenamefont {Abbott}\ \emph {et~al.}(2024)\citenamefont {Abbott} \emph {et~al.}}]{DES:2024jxu}%
  \BibitemOpen
  \bibfield  {author} {\bibinfo {author} {\bibfnamefont {T.~M.~C.}\ \bibnamefont {Abbott}} \emph {et~al.} (\bibinfo {collaboration} {DES}),\ }\href {\doibase 10.3847/2041-8213/ad6f9f} {\bibfield  {journal} {\bibinfo  {journal} {Astrophys. J. Lett.}\ }\textbf {\bibinfo {volume} {973}},\ \bibinfo {pages} {L14} (\bibinfo {year} {2024})},\ \Eprint {http://arxiv.org/abs/2401.02929} {arXiv:2401.02929 [astro-ph.CO]} \BibitemShut {NoStop}%
\bibitem [{\citenamefont {Popovic}\ \emph {et~al.}(2025)\citenamefont {Popovic} \emph {et~al.}}]{Popovic:2025glk}%
  \BibitemOpen
  \bibfield  {author} {\bibinfo {author} {\bibfnamefont {B.}~\bibnamefont {Popovic}} \emph {et~al.},\ }\href@noop {} {\  (\bibinfo {year} {2025})},\ \Eprint {http://arxiv.org/abs/2506.05471} {arXiv:2506.05471 [astro-ph.CO]} \BibitemShut {NoStop}%
\bibitem [{\citenamefont {Popovic}\ \emph {et~al.}(2026)\citenamefont {Popovic} \emph {et~al.}}]{DES:2025sig}%
  \BibitemOpen
  \bibfield  {author} {\bibinfo {author} {\bibfnamefont {B.}~\bibnamefont {Popovic}} \emph {et~al.} (\bibinfo {collaboration} {DES}),\ }\href {\doibase 10.1093/mnras/stag632} {\bibfield  {journal} {\bibinfo  {journal} {Mon. Not. Roy. Astron. Soc.}\ }\textbf {\bibinfo {volume} {548}},\ \bibinfo {pages} {stag632} (\bibinfo {year} {2026})},\ \Eprint {http://arxiv.org/abs/2511.07517} {arXiv:2511.07517 [astro-ph.CO]} \BibitemShut {NoStop}%
\bibitem [{\citenamefont {Scolnic}\ \emph {et~al.}(2018)\citenamefont {Scolnic} \emph {et~al.}}]{Pan-STARRS1:2017jku}%
  \BibitemOpen
  \bibfield  {author} {\bibinfo {author} {\bibfnamefont {D.~M.}\ \bibnamefont {Scolnic}} \emph {et~al.} (\bibinfo {collaboration} {Pan-STARRS1}),\ }\href {\doibase 10.3847/1538-4357/aab9bb} {\bibfield  {journal} {\bibinfo  {journal} {Astrophys. J.}\ }\textbf {\bibinfo {volume} {859}},\ \bibinfo {pages} {101} (\bibinfo {year} {2018})},\ \Eprint {http://arxiv.org/abs/1710.00845} {arXiv:1710.00845 [astro-ph.CO]} \BibitemShut {NoStop}%
\bibitem [{\citenamefont {Moresco}\ \emph {et~al.}(2012)\citenamefont {Moresco} \emph {et~al.}}]{Moresco:2012jh}%
  \BibitemOpen
  \bibfield  {author} {\bibinfo {author} {\bibfnamefont {M.}~\bibnamefont {Moresco}} \emph {et~al.},\ }\href {\doibase 10.1088/1475-7516/2012/08/006} {\bibfield  {journal} {\bibinfo  {journal} {JCAP}\ }\textbf {\bibinfo {volume} {08}},\ \bibinfo {pages} {006} (\bibinfo {year} {2012})},\ \Eprint {http://arxiv.org/abs/1201.3609} {arXiv:1201.3609 [astro-ph.CO]} \BibitemShut {NoStop}%
\bibitem [{\citenamefont {{Akaike}}(1974)}]{1974ITAC...19..716A}%
  \BibitemOpen
  \bibfield  {author} {\bibinfo {author} {\bibfnamefont {H.}~\bibnamefont {{Akaike}}},\ }\href {\doibase 10.1109/TAC.1974.1100705} {\bibfield  {journal} {\bibinfo  {journal} {IEEE Transactions on Automatic Control}\ }\textbf {\bibinfo {volume} {19}},\ \bibinfo {pages} {716} (\bibinfo {year} {1974})}\BibitemShut {NoStop}%
\bibitem [{\citenamefont {Schwarz}(1978)}]{Schwarz:1978tpv}%
  \BibitemOpen
  \bibfield  {author} {\bibinfo {author} {\bibfnamefont {G.}~\bibnamefont {Schwarz}},\ }\href@noop {} {\bibfield  {journal} {\bibinfo  {journal} {Annals Statist.}\ }\textbf {\bibinfo {volume} {6}},\ \bibinfo {pages} {461} (\bibinfo {year} {1978})}\BibitemShut {NoStop}%
\bibitem [{\citenamefont {Trotta}(2008)}]{Trotta:2008qt}%
  \BibitemOpen
  \bibfield  {author} {\bibinfo {author} {\bibfnamefont {R.}~\bibnamefont {Trotta}},\ }\href {\doibase 10.1080/00107510802066753} {\bibfield  {journal} {\bibinfo  {journal} {Contemp. Phys.}\ }\textbf {\bibinfo {volume} {49}},\ \bibinfo {pages} {71} (\bibinfo {year} {2008})},\ \Eprint {http://arxiv.org/abs/0803.4089} {arXiv:0803.4089 [astro-ph]} \BibitemShut {NoStop}%
\end{thebibliography}%
\end{document}